\documentclass[final,times,onecolumn]{elsarticle}
\journal{Physica A}
\usepackage{subfig}
\usepackage{amsmath}
\usepackage{color}

\begin{document}

\begin{frontmatter}
\title{Estimation of connectivity measures in gappy time series}
\author[label1]{G. Papadopoulos}
\author[label2]{D. Kugiumtzis}\ead{dkugiu@gen.auth.gr}
\address[label1]{Economics Department, Democritus University of Thrace, Komotini, Greece}
\address[label2]{Department of Electrical and Computer Engineering, Faculty of Engineering, Aristotle University of Thessaloniki, 54124 Thessaloniki, Greece}

\begin{abstract}
A new method is proposed to compute connectivity measures on
multivariate time series with gaps. Rather than removing or
filling the gaps, the rows of the joint data matrix containing
empty entries are removed and the calculations are done on the
remainder matrix. The method, called measure adapted gap removal
(MAGR), can be applied to any connectivity measure that uses a
joint data matrix, such as cross correlation, cross mutual
information and transfer entropy. MAGR is favorably compared using
these three measures to a number of known gap-filling techniques,
as well as the gap closure. The superiority of MAGR is 
illustrated on time series from synthetic systems and financial time
series.
\end{abstract}

\end{frontmatter}

\section{Introduction}
\label{sec:Introduction}
\par
In the analysis of multivariate time series, the primary interest
is in investigating interactions among the observed variables. For
this a number of measures have been proposed under different
terms, such as inter-dependence, coupling, Granger causality and
connectivity. There are certain distinctions of these measures, as
correlation and causality measures, linear and nonlinear measures,
and measures on the time and frequency domain
\cite{Hlavackova07,Dahlhaus08,Bressler11,Papana13b}. Examples of
such measures that we use in our study are the correlation
measures of cross correlation and cross mutual information, and
the causality measure of transfer entropy \cite{Schreiber00}.

\par
All these methods are developed under the assumption that the time
series being analyzed are evenly spaced, meaning the measurements
are taken at a fixed sampling rate. However, this is not always
the case and in many applications the time series have gaps, as in
environmental sciences (occurrence of gaps is a common problem
with geophysical \cite{Kondrashov06,Dergachev01,Elshorbagy02},
ecological \cite{Facchini11}, and oceanographic
\cite{Ustoorikar08} time series), astronomy \cite{Kondrashov10}
and socio-economics \cite{Harvey84,Warga92}. Sampling at irregular
or uneven time intervals regards a different class of problems
and it is not studied here, e.g. for spectral estimation see
\cite{Broersen09,Hocke09} and for Granger causality see
\cite{Bahadori12}.

\par
The common approach of all proposed techniques for gappy time
series is first to fill the gaps in some way and then apply the
method of choice to the new evenly spaced time series. The
techniques range from relatively simple ones, such as the ``gap
closure'' joining the edges of the gaps, cubic spline and
$k$-nearest neighbors interpolation, to more complex ones such as
Single Spectrum Analysis (SSA) \cite{Kondrashov06}, neural
networks \cite{Dergachev01}, and state space reconstruction under
the hypothesis of chaos \cite{Paparella05}, among others.
Comparisons of these methods on different real-world applications
can be found in \cite{Elshorbagy02,Kreindler06,Musial11}.

\par
For bivariate and multivariate time series, methods such as SSA
and neural networks can be extended to incorporate information
from all time series to recover the gaps \cite{Kondrashov06},
along with other recent approaches making use of the concepts of
nonlinear dynamics and surrogate data \cite{Facchini11}.
Especially for the application of transfer entropy, Kulp and Tracy
\cite{Kulp09} examined a stochastic gap-filling technique called
``random replacement'' in gappy data from harmonic oscillators.

\par
In our paper we take a different route and address the problem in
a method specific manner. Instead of filling the gappy time
series, we modify the measure to be used, accounting for the gaps
in the time series, thus leaving the underlying dynamics to the
time series intact, free of artificial intervention. Our approach,
called measure adapted gap removal (MAGR), is general for any
measure of multivariate time series, and we exemplify it here on
two correlation measures, cross correlation and cross mutual
information, and one Granger causality measure, the transfer
entropy. We demonstrate the effectiveness of our approach in
comparison to gap-filling methods on a linear stochastic
multivariate autoregressive (MVAR) system and a nonlinear system,
the coupled Henon map. We randomly remove samples of the generated
time series and estimate each measure on the gappy time series
using our approach as well as different gap filling methods.
Further, we consider also the case of missing blocks of
consecutive samples, of fixed or varying size, often met in
applications.

\par
The remainder of the paper is structured as follows.
Section~\ref{sec:Measures} gives briefly the theoretical framework
of the correlation and causality measures used in this study.
Section~\ref{sec:MAGR} describes our approach for computing the
measures on gappy time series. Section~\ref{sec:Simulations}
presents the simulation results on the linear and nonlinear
systems for the estimation of the measures with our approach and
the gap-filling methods. Section~\ref{sec:Application} presents an 
application of MAGR to real financial data. Finally, the results are discussed in
Section~\ref{sec:Discussion}.

\section{Correlation and causality measures}
\label{sec:Measures}
\par
In the following, we present briefly the three measures used to
demonstrate our approach when the multivariate time series contain
single missing values, called single gaps, or blocks of missing
values, called block gaps. We denote the variables with capital
letters and the sample values with small letters. The measures
considered in this study are bivariate and for multivariate time
series they are applied to each pair of time series. The
implementation of our approach to multivariate measures, e.g. the
partial transfer entropy \cite{Vakorin09,Papana12}, is
straightforward.

\subsection{Cross correlation}
\label{subsec:cc}
\par
For two simultaneously measured variables $X$ and $Y$ giving the
time series $\{x_t,y_t\}_{t=1}^N$, the cross correlation measures the
linear correlation of $X$ and $Y$ at the same time $t$, or at a
delay $\tau$, defined as
\begin{equation}
r_{XY}(\tau) = \mbox{Corr}(X_t,Y_{t+\tau})=\frac{\sum_{t=1}^{N-\tau} (x_t-\bar{x})(y_{t+\tau}-\bar{y})}{\sqrt{ \sum_{t=1}^{N-\tau} (x_t-\bar{x})^2 \sum_{t=1}^{N-\tau} (y_t-\bar{y})^2}},
\label{eq:cc}
\end{equation}
where $\bar{x}$ and $\bar{y}$ are the mean values of the two time series. For $\tau=0$, $r_{XY}(0)$ is the standard Pearson correlation coefficient of $X$ and $Y$.

\subsection{Cross mutual information}
\label{subsec:cmi}
\par
Cross mutual information is an appropriate analogue to cross
correlation if also nonlinear correlation is to be estimated.
First, mutual information of two variables $X$ and $Y$ is defined
in terms of entropies as
\begin{equation}
I(X;Y) = H(X) + H(Y) - H(X,Y),
\label{eq:MI1}
\end{equation}
where $H(X)$ and $H(X,Y)$ are the Shannon entropy of $X$ and the
joint entropy of $(X,Y)$, respectively \cite{Cover91}. For the
estimation of the entropies, we first discretize $X$ and $Y$, and
then compute the standard frequency estimates of the probability
mass function of $X$, $Y$ and $(X,Y)$ denoted $p_X$, $p_Y$ and
$p_{X,Y}$, respectively, giving
\begin{equation}
I(X;Y) = \sum_{x} \sum_{y} p_{X,Y}(x,y) \log \frac{p_{X,Y}(x,y)}{p_{X}(x) p_{Y}(y)}.
\label{eq:MI2}
\end{equation}

\par
We consider here the discretization of $X$ and $Y$ using
equiprobable partition of their domains in $b$ intervals. So,
there is an equal occupancy at each interval, and the probability
$p_{X}(x)$ of the $x$-th element of the partition of $X$ is
$p_{X}(x)=1/b$ and respectively $p_{Y}(y)=1/b$ for $Y$. The
estimation of the mutual information depends on the selected
number of bins $b$, and here we use $b=\sqrt{N/5}$
\cite{Cellucci05,Papana09}.

\par
The cross mutual information for the time series
$\{x_t,y_t\}_{t=1}^N$ is simply the mutual information of $X_t$
and $Y_{t+\tau}$, $I_{XY}(\tau)=I(X_t,Y_{t+\tau})$. Note that when
$\tau=0$ the two measures $r_{XY}(0)$ and $I_{XY}(0)$ are
symmetrical and indicate the correlation of $X_t$ and $Y_t$, while
for $\tau>0$, a significant value of $r_{XY}(\tau)$ or
$I_{XY}(\tau)$ indicates correlation of $X_t$ and $Y_{t+\tau}$,
which can be interpreted loosely as a drive-response relationship
from $X$ to $Y$ \cite{Kullmann02}. For the latter, more
appropriate measures have been developed, termed as Granger
causality measures.

\subsection{Transfer entropy}
\label{subsec:te}
\par
A nonlinear measure of Granger causality from information theory
is the transfer entropy \cite{Schreiber00}. Transfer entropy
quantifies the information flow from a system represented by a
variable $X$ to another system represented by a variable $Y$ at a
leading time (originally taken to be one time step ahead), and it
can thus be regarded as a Granger causality measure. The transfer
entropy $\mbox{TE}_{X \rightarrow Y}$ is actually the conditional
mutual information $I(Y_{t+1};\textbf{X}_t| \textbf{Y}_t)$, where
$\textbf{X}_t=[X_t,X_{t-\tau},\ldots,X_{t-(m-1)\tau}]^{\mbox{\scriptsize
T}}$ is the reconstructed vector variable for $X$ and
$\textbf{Y}_t=[Y_t,Y_{t-\tau},\ldots,Y_{t-(m-1)\tau}]^{\mbox{\scriptsize
T}}$ for $Y$, and represents the dependence of $Y_{t+1}$ on the
history of $X$ accounting for its own history. The parameter $m$
is the embedding dimension and $\tau$ is the delay time, taken to
be the same for both $X$ and $Y$ as this choice was found optimal
for the detection of coupling \cite{Papana11}. The conditional
mutual information can be expressed in terms of entropies and we
have
\begin{equation}
\mbox{TE}_{X \rightarrow Y} =    %
- H(Y_{t+1},\textbf{X}_t,\textbf{Y}_t) + H(\textbf{X}_t,\textbf{Y}_t) %
+ H(Y_{t+1},\textbf{Y}_t) - H(\textbf{Y}_t).                         %
\label{eq:TE1}
\end{equation}
\par
The binning estimate of the entropies of high-dimensional vector variables is not appropriate and instead we use the entropy estimate given by the correlation sum, $H(\textbf{X}_t) = \log C(\textbf{X}_t,r) + \log r$ \cite{Manzan02}, where the correlation sum for a distance $r$ is
\begin{equation}
 C(\textbf{X}_t,r) = \frac{2}{(N-(m-1)\tau)(N-(m-1)\tau+1)} \sum_{i=(m-1)\tau+1}^N \sum_{j=i+1}^N \Theta(r-\|\textbf{x}_i-\textbf{x}_j\|),
\label{eq:CS}
\end{equation}
where $\Theta(x)$ is the Heaviside function being one if $x>0$ and
zero otherwise, and $\textbf{x}_i$ is the sample reconstructed
vector of $X$ at time step $i$. Substituting the estimation of
entropies in terms of correlation sums gives \cite{Papana11}
\begin{equation}
\mbox{TE}_{X \rightarrow Y} =                                   %
\log{\frac{C(Y_{t+1},\textbf{X}_t,\textbf{Y}_t)C(\textbf{Y}_t)} %
{C(\textbf{X}_t,\textbf{Y}_t)C(Y_{t+1},\textbf{Y}_t)}}.         %
\label{eq:TE2}
\end{equation}

\par
The estimation of transfer entropy making use of correlation sums
depends on the distance $r$ similarly to the dependence of the
binning estimate on the number of bins $b$. The selection of $r$
varies in applications using the correlation sum and one strategy
is to search for the optimal $r$ in a predefined range. We
followed this strategy for the example with the uni-directionally
coupled Henon map for embedding dimensions $m = 1$ and $m = 2$
($\tau = 1$). The optimal $r$ giving maximum discrimination 
in the two directions (highest value of $\mbox{TE}_{X\rightarrow Y}$ and 
closest to zero $\mbox{TE}_{Y\rightarrow X}$) is found to be $r=0.2$, where the time series
were normalized to zero mean and one standard deviation. Thus in
the study we set $r=0.2$.

\section{Measure adapted gap removal}
\label{sec:MAGR}
\par
The existing methodology for time series with gaps regards
methods that attempt to fill the gaps in the time series and then
proceed with the analysis of the derived time series. Any such
approach fills the missing values using a model under some
assumption about the missing values, intervening in this way
artificially -- or even arbitrarily if there are no grounds for
the particular model choice -- to the underlying dynamics. The
most arbitrary approach is the gap closure (GC) suppressing the
gaps in the time series and connecting the edges, which is valid
only when there are no dependencies in the time series. We follow
a different strategy, leave the gaps in the time series, and instead
adjust each method of analysis accounting for the gaps. We call this
approach measure adapted gap removal (MAGR). In this study we
choose the three connectivity measures of multivariate time series
to demonstrate our approach, but the approach can be extended to
many other measures. Actually, it can be applied to any method of
multivariate time series analysis that makes use of temporally
close or sequentially ordered samples of the time series. The
rationale is simply to use only the sample sets that do not contain
gaps. We illustrate below the MAGR approach in detail.

\par
First, for measures that do not require reconstructed vectors for
their estimation, such as the linear cross-correlation and the
cross mutual information, the problem can be addressed quite
easily: we take the intersection of the non-empty values in the
pairs $(x_t,y_{t+\tau})$. To illustrate this, let us consider two
time series for the variables $X$ and $Y$ with gaps, as shown in
Table~\ref{tab:example1} ($N=10$).
\begin{table}
\caption{Two time series $X$ and $Y$ with gaps (columns 2 and 3),
where the time series $Y$ is also displaced by one time step ahead
(column 4) and back (column 6) and $X$ one time step back (column
5).} \centerline{\begin{tabular}{cccccc} \hline
time & $X_t$ & $Y_t$ & $Y_{t+1}$ & $X_{t-1}$ & $Y_{t-1}$ \\ \hline %
1 & $x_1$ & $y_1$ & $y_2$ & & \\
2 & $x_2$ & $y_2$ & $y_3$ & $x_1$ & $y_1$ \\
3 & $x_3$ & $y_3$ & -- & $x_2$ & $y_2$  \\
4 & $x_4$ & -- & $y_5$ & $x_3$ & $y_3$ \\
5 & -- & $y_5$ & $y_6$ & $x_4$ & -- \\
6 & $x_6$ & $y_6$ & $y_7$ & -- & $y_5$ \\
7 & $x_7$ & $y_7$ & $y_8$ & $x_6$ & $y_6$ \\
8 & -- & $y_8$ & $y_9$ & $x_7$ & $y_7$ \\
9 & $x_9$ & $y_9$ & $y_{10}$ & -- & $y_8$ \\
10 & $x_{10}$ & $y_{10}$ & & $x_9$ & $y_9$ \\ \hline %
\end{tabular}}
\label{tab:example1}
\end{table}
Examining the paired values in columns 2 and 3 of
Table~\ref{tab:example1} for empty entries, we leave out the pairs
for time steps 4, 5 and 8, and make the computations for
$r_{XY}(0)$ and $I_{XY}(0)$ in (\ref{eq:cc}) and (\ref{eq:MI2}),
respectively, on the remaining pairs of values. For non-zero lags,
the same technique applies but now one time series is time-shifted
by the given time lag. Assuming time lag one, $r_{XY}(1)$ and
$I_{XY}(1)$ can be computed using the pairs $(x_t,y_{t+1})$ of
non-empty values, and for the example in Table~\ref{tab:example1}
these are for time steps 1,2,4,6,7,9.

\par
This approach gets more complicated and more data points are
discarded when the connectivity measure requires state space
reconstruction of time series, such as the measure of transfer
entropy. For $\mbox{TE}_{X \rightarrow Y}$, the requirement for
non-empty values for each time step $t$ involves $y_{t+1}$ as well
as
$\textbf{x}_t=[x_t,x_{t-\tau},\ldots,x_{t-(m-1)\tau}]^{\mbox{\scriptsize
T}}$ and
$\textbf{y}_t=[y_t,y_{t-\tau},\ldots,y_{t-(m-1)\tau}]^{\mbox{\scriptsize
T}}$. For the simplest case of $m = 1$ and $\tau = 1$, the
computation of TE is based on the joint data matrix consisting of
triplets $(y_{t+1},x_t,y_t)$ for each time step $t$ (row in the
matrix), and each triplet has to be checked for empty values. For
the example in Table~\ref{tab:example1}, more data points have to
be discarded, for time steps 3,4,5 and 8. As embedding dimension
$m$ increases, the number of discarded data points increases as
well. Denoting the number of gaps in $X$ as $g_X$ and in $Y$ as
$g_Y$, respectively, if only $X$ has gaps then the maximum number
of discarded data points is $d=m g_X$, and respectively if only
$Y$ has gaps $d=(m+1) g_Y$, while for both $X$ and $Y$ having gaps
we have $d=m g_X +(m+1) g_Y$. The number of discarded data points
is smaller than $d$ if the gaps occur on consecutive time steps or
synchronously in $X$ and $Y$. For example, the computation of
$\mbox{TE}_{X \rightarrow Y}$ for $m=2$ and $\tau=1$ requires the
pentad $(y_{t+1},x_t,x_{t-1},y_t,y_{t-1})$ having non-empty
values, and considering the time series of
Table~\ref{tab:example1}, there are only two such pentads for time
steps 2 and 7.

\section{Simulations and Results}
\label{sec:Simulations}
\par

\subsection{Simulation setup}
\label{subsec:setup}
\par
To examine the effectiveness of MAGR we generate time series from a linear stochastic and a chaotic system, where a predefined number of samples are randomly removed, and compare the results of measure estimation obtained using MAGR to these of other gap-filling algorithms. In the simulation study we consider the standard gap-filling techniques using linear interpolation (LI), cubic interpolation (CI), splines interpolation (SPI), nearest neighbor interpolation (NNI) and the stochastic interpolation (STI) proposed in \cite{Kulp09}, as well as the gap closure technique (GC).

\par
The first system is the linear multivariate autoregressive process (MVAR)
\begin{equation}
\begin{array}{l}
X_t = 1.2 X_{t-1} - 0.95 X_{t-2} +W_t^X \\
Y_t = -0.5 X_{t-1} - 0.4 Y_{t-9} +W_t^Y \\
\end{array}
\label{eq:system1}
\end{equation}
where $W_t^X$ and $W_t^Y$ are white noise processes (mean zero and
standard deviation one) uncorrelated to each other (the system is
actually formed by the two first equations of the system in
\cite{Wehling08}). The second system is the uni-directionally
coupled Henon map
\begin{equation}
\begin{array}{l}
X_{t+1} = 1.4  - X^{2}_t + 0.3X_{t-1} \\
Y_{t+1} = 1.4 - C X_t Y_t - (1-C)Y^2_t + 0.3Y_{t-1} \\
\end{array}
\label{eq:system2}
\end{equation}
where $C$ is the coupling strength parameter
\cite{QuianQuiroga00}. For linear cross-correlation and cross
mutual information measures, we set $C=0.7$ close to complete
synchronization, while for transfer entropy estimation we set
$C=0.4$ to have moderate coupling for which $\mbox{TE}_{X
\rightarrow Y}$ obtains its maximum and can thus detect best the
causality effect $X \rightarrow Y$. Note that for any $C$ we
expect to have $\mbox{TE}_{Y \rightarrow X} \simeq 0$.

\par
For the evaluation of MAGR and the other methods treating the
gaps, filling or closing them, we compute the correlation and
causality measures on the time series without gaps and on the time
series with gaps. For the latter the gaps are first filled or
removed by each of the techniques GC, LI, CI, SPI, NNI and STI,
and then the connectivity measure is computed, while for MAGR the
connectivity measure is computed after removing the rows of the
joint data matrix containing missing values. The performance of
the gap treatment is thus evaluated by the difference of the two
estimates for no gaps and gaps. For each simulation, we generate
two time series of length $N$ for $X$ and $Y$, termed as the
original time series. From each of the time series we randomly
remove $g$ samples ($g=g_X=g_Y$ with the gaps being separately
selected for each time series), ranging from 5\% to 50\% of $N$.
Using each gap-filling algorithm we fill the gaps and obtain new
time series of length $N$, while using the gap closure the
obtained time series have length $N-g$. The correlation and
causality measures are estimated on the gap-filled or gap-closed
time series and also on the original time series of respective
length, i.e. $N$ for the gap-filling algorithms and $N-g$ for the
gap-closure. The reason for matching the lengths is to suppress
the effect of the time series length in the estimation of the
measure, so as to compare directly the measure results and assess
the performance of the gap treatment. To achieve direct comparison
for MAGR, for each correlation or causality measure we obtain
first the corresponding data matrix after removing the rows
containing empty entries (see Table~\ref{tab:example1}). Depending
on the number of these rows $N_r$, the matching length of the
original time series is $N-N_r$. Note that this length may change
for each measure and parameter value.

In order to assess the effectiveness of each method we take the
difference of the correlation or causality measure on the original
time series from that of the gap treated time series of matching
length, denoted as $\mbox{d}r$, $\mbox{d}I$ and dTE (the
corresponding variable indices are not shown here). Finally, the
above calculations are repeated 50 times, to draw statistically
safe conclusions, and the mean values are reported.

\subsection{Linear cross-correlation and cross mutual information estimation}
\label{subsec:resultsccmi}
\par

\begin{figure}[htb]
\hbox{\centerline{\includegraphics[width=7cm]{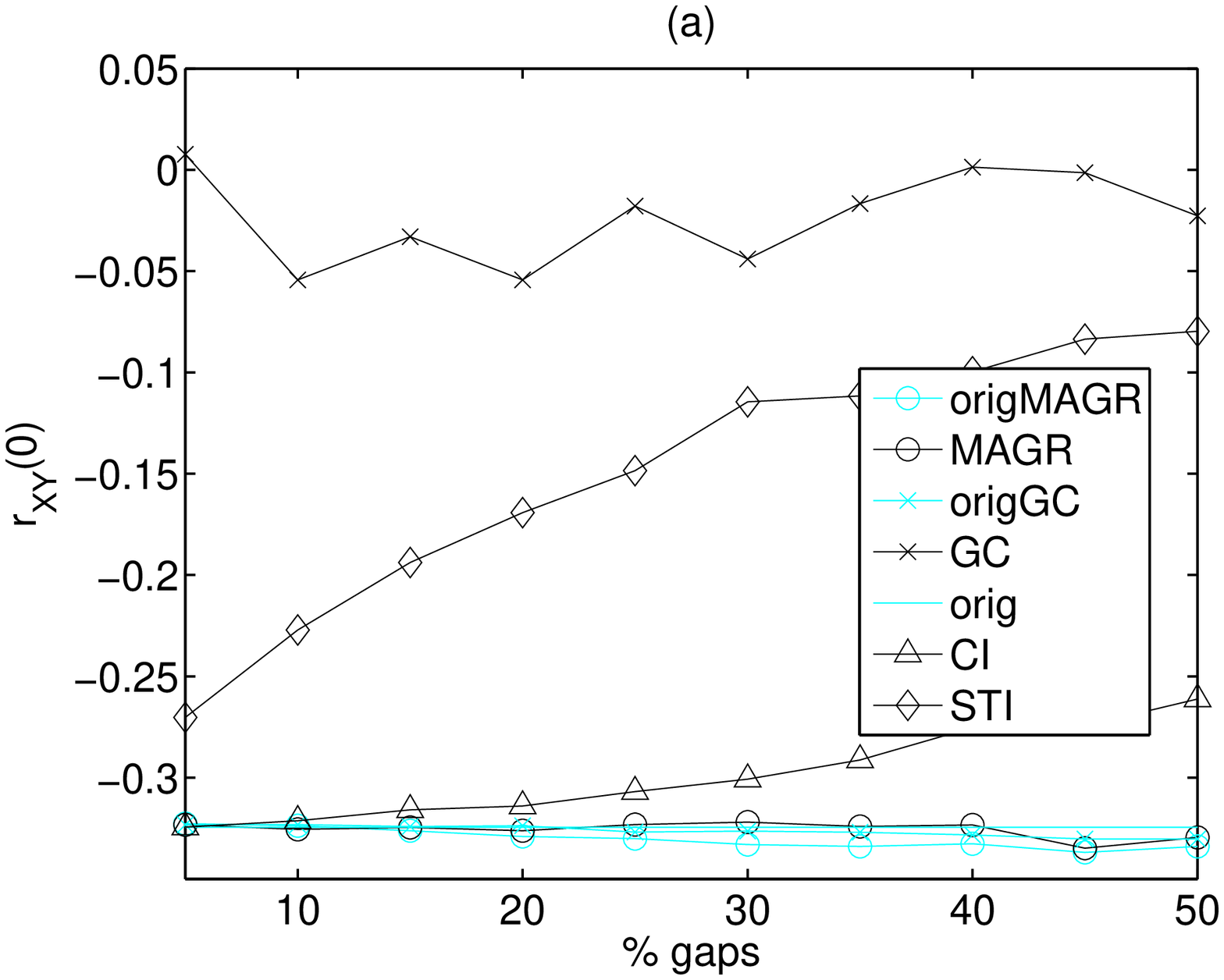}
\includegraphics[width=7cm]{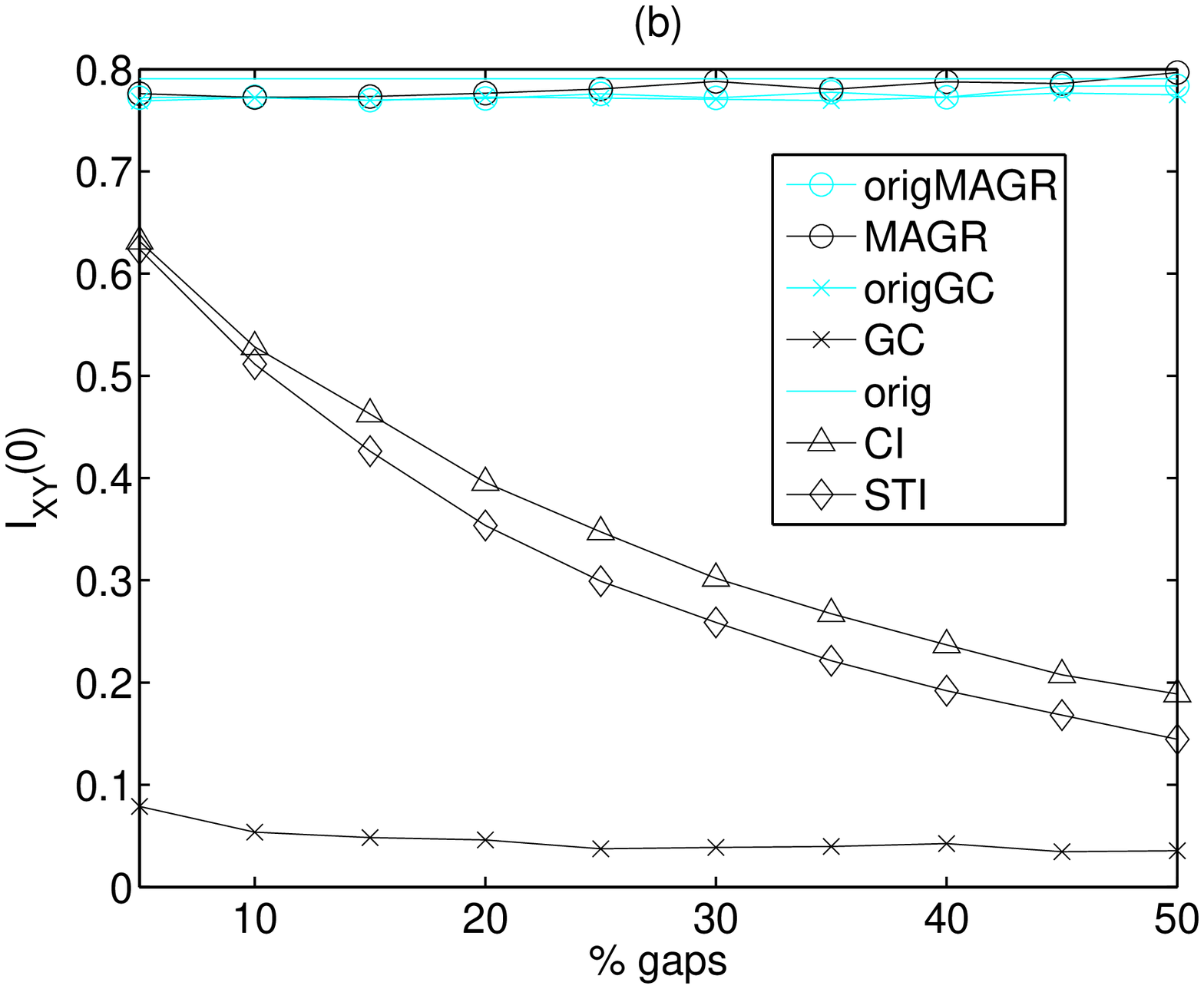}}}
\caption{Zero lag cross correlation for the MVAR system in (a) and
zero lag cross mutual information for the coupled Henon map in (b)
as a function of the percentage of gaps in time series of length
$N=500$, using MAGR, gap closure (GC), cubic interpolation (CI)
and stochastic interpolation (STI). Superimposed are also shown
the estimates for time series without gaps of appropriate matching
length for each gap treating method (denoted by 'orig' in the
legend).} \label{fig1}
\end{figure}

First we note that the methods for filling or removing gaps give
very different estimates of correlation. For example, as shown in
Fig.~\ref{fig1}a, while for the original non-gappy time series of
$N=500$ from the MVAR system we have $r_{XY}(0)=-0.325$, the
gap-filling approaches cannot match this value, but $r_{XY}(0)$
deviates towards zero as $g$ increases (STI failing more than CI).
Gap-closure (GC) performs worst, giving $r_{XY}(0)$ at the zero
level for any $g$, because even if there is a single gap in the
time series, after gap closing the subsequent sample pairs are not
matched. This failure of GC is constantly observed in all the
simulations. On the other hand, MAGR stays at the level of the
original $r_{XY}(0)$ for any $g$, up to 50\% of $N$ that we have
tested. The same applies for cross mutual information. An
illustrative example is shown in Fig.~\ref{fig1}b for $I_{XY}(0)$
and the coupled Henon system. The only notable difference to
Fig.~\ref{fig1}a, is that CI and STI follow the same pattern of
decrease of $I_{XY}(0)$ towards the zero level with $g$.

\begin{figure}[htb]
\hbox{\centerline{\includegraphics[width=7cm]{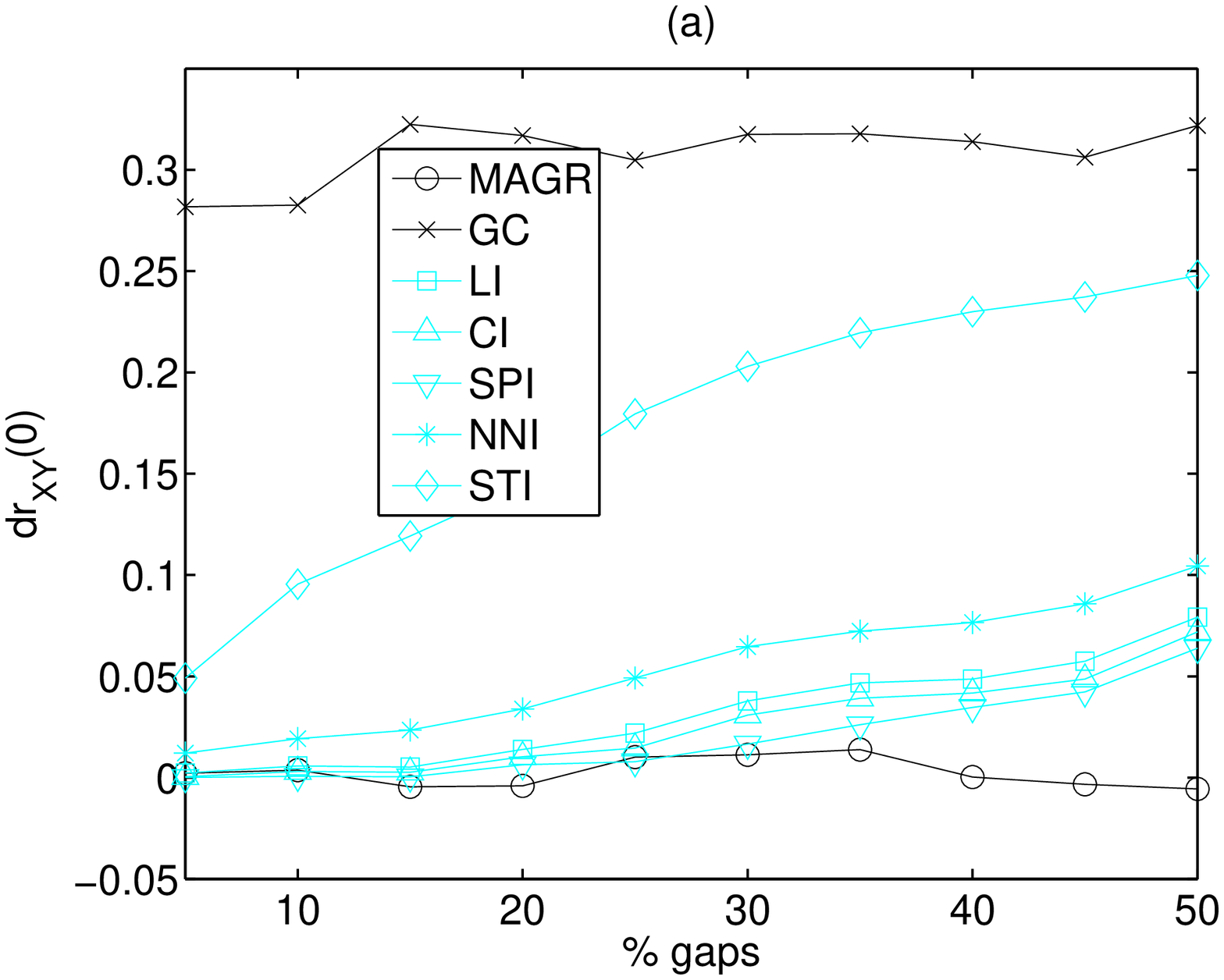}
\includegraphics[width=7cm]{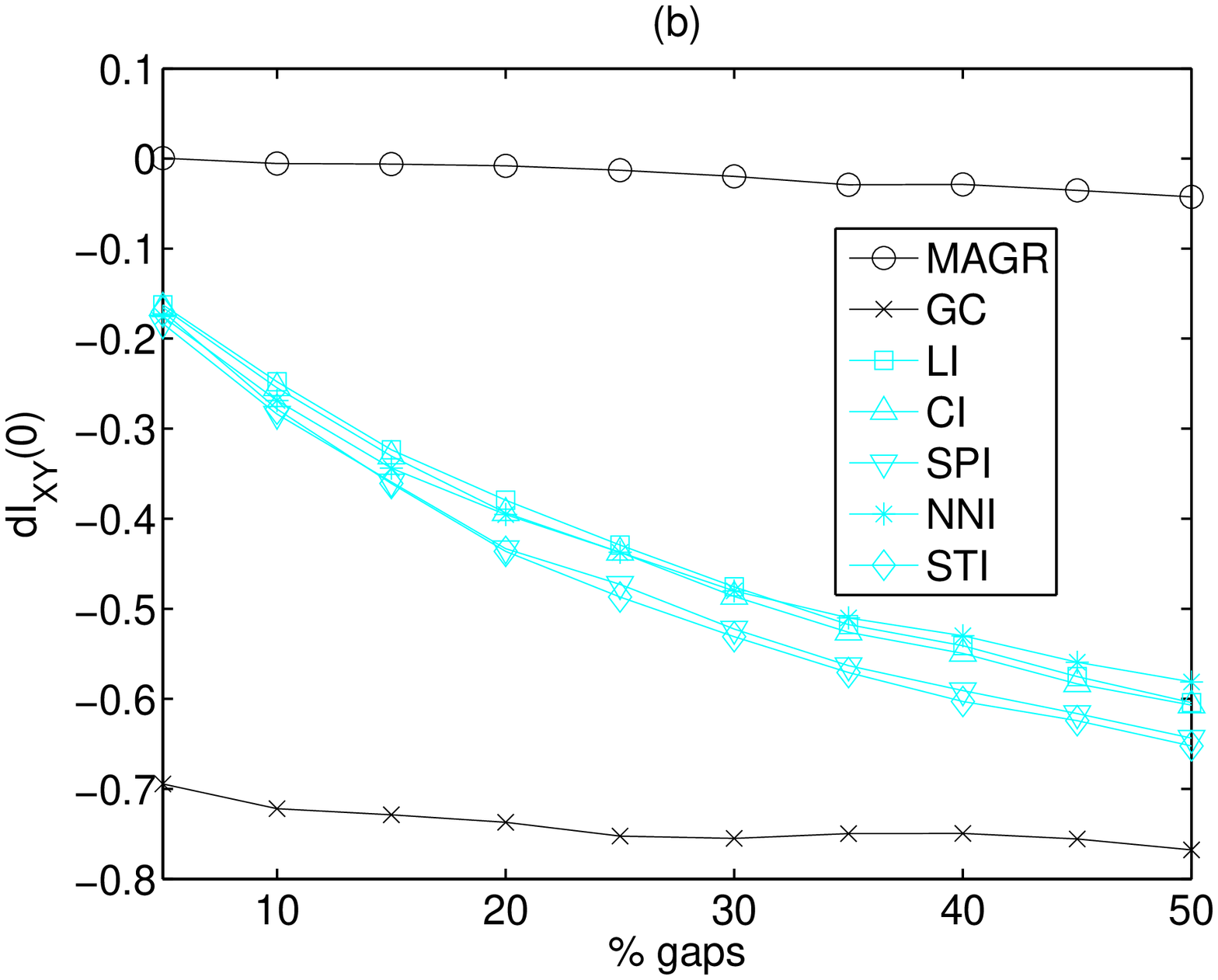}}}
\caption{Difference in the zero lag cross correlation, d$r_{XY}(0)$, for the MVAR system in (a) and zero lag cross mutual information, d$I_{XY}(0)$, for the coupled Henon map in (b) as a function of the percentage of gaps in time series of length $N=500$, for the gap-treating techniques as shown in the legend.}
\label{fig2}
\end{figure}

The effectiveness of gap treating techniques can be better
expressed by the performance of d$r$ or d$I$ that should optimally
be at the zero level. In Fig.~\ref{fig2}, d$r$ and d$I$ are shown
for the same experimental setup as for Fig.~\ref{fig1}, adding
also the results for three other gap-filling methods.
The bound of worst performance is set by GC giving the largest
difference d$r$ or d$I$ (the correlation estimate of the
gap-closured time series is at the zero level), while MAGR
actually lies at the bound of best performance, giving d$r$ or
d$I$ at the zero level (the correlation estimate does not change
after applying MAGR). In-between these two bounds lies the
performance of the gap-filling techniques at a varying order but
following the same pattern of d$r$ or d$I$ deviating from zero
with $g$. For the linear system in Fig.~\ref{fig2}a, all but STI
gap-filling techniques perform quite adequately for small $g$ and
gradually worsen for larger $g$. STI performs worse, possibly
because it fills the gaps essentially with random numbers, while
the other gap-filling techniques assume some simple relationship
of the edge points of the gaps that here seems to be better than
random. This seems to work satisfactorily when there are few gaps
in the time series, say up to 25\% of $N$ for the linear system.
However, when the structure of the time series is nonlinear, all
gap-filling techniques fail in the same way, as shown in
Fig.~\ref{fig2}b.

\subsection{Transfer entropy estimation}
\label{subsec:resultste}
\par

\begin{figure}[htb]
\hbox{\centerline{\includegraphics[width=7cm]{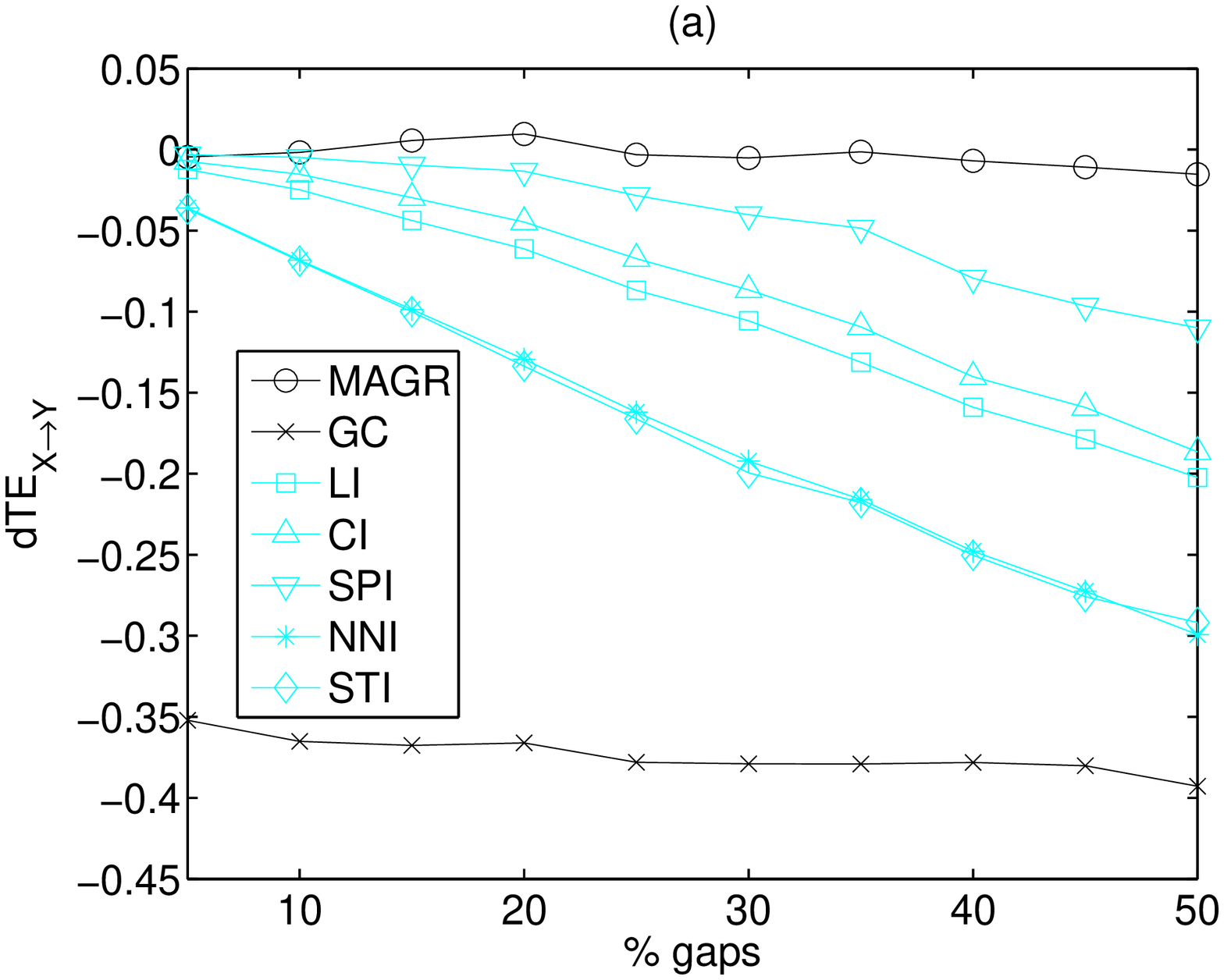}
\includegraphics[width=7cm]{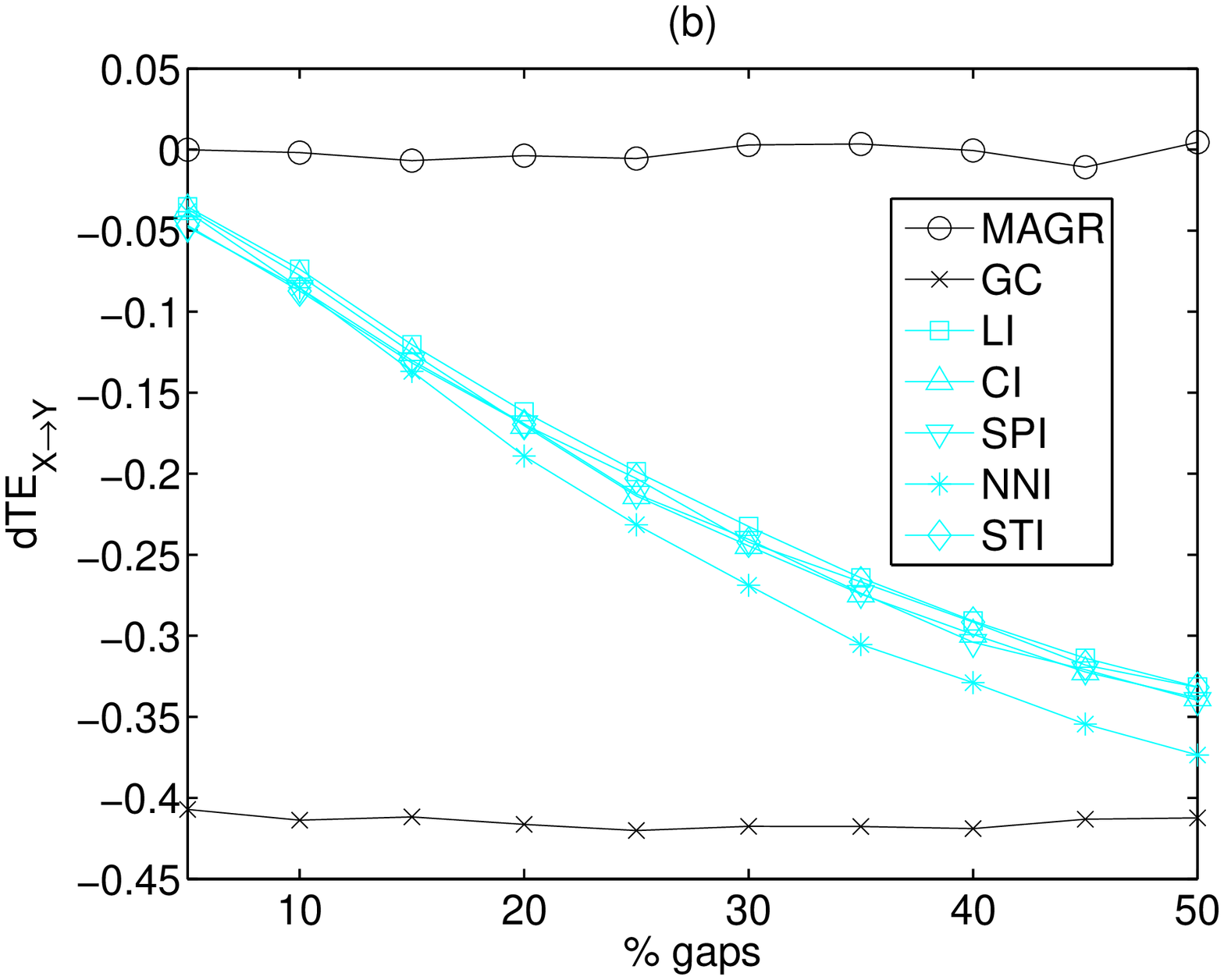}}}
\hbox{\centerline{\includegraphics[width=7cm]{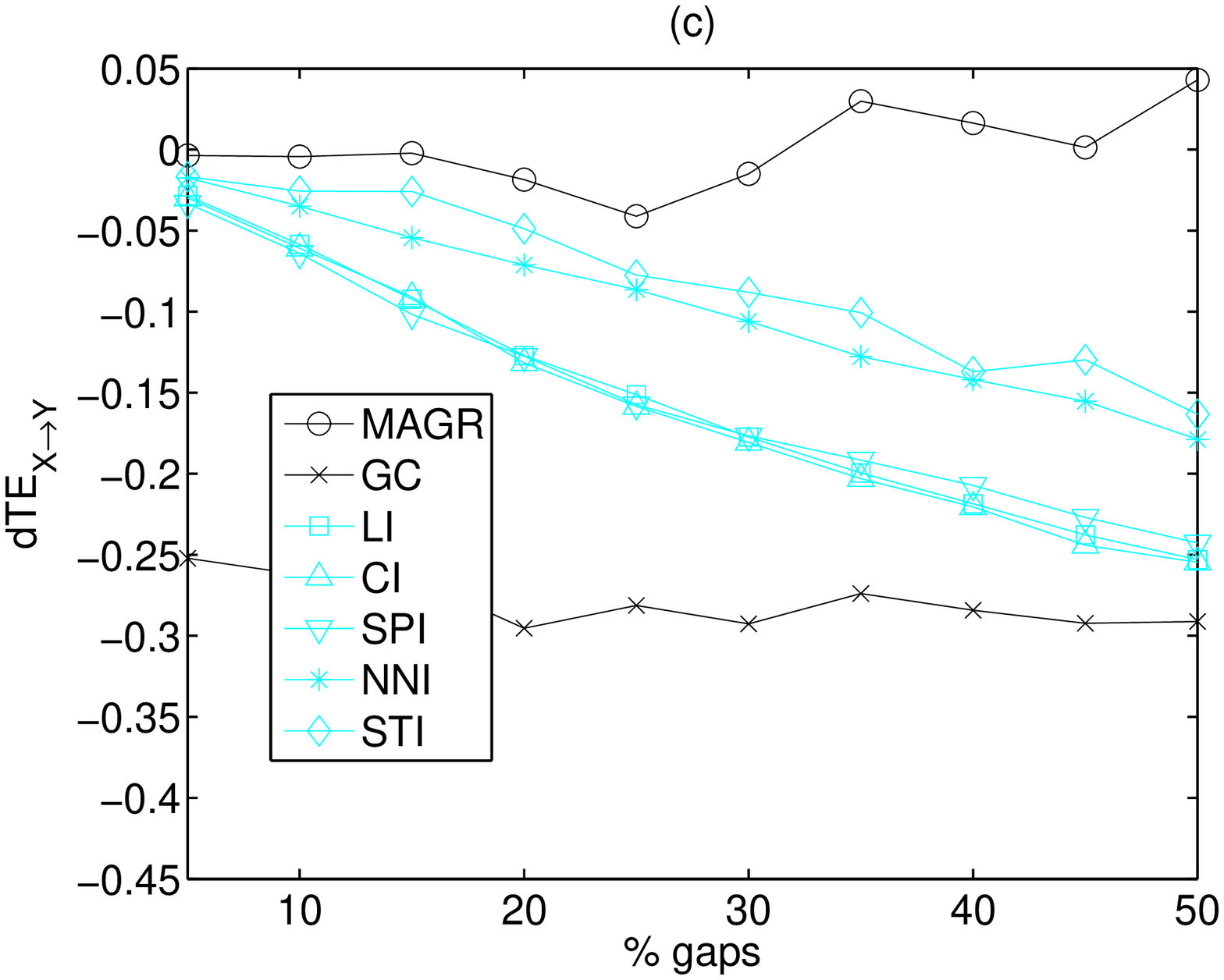}
\includegraphics[width=7cm]{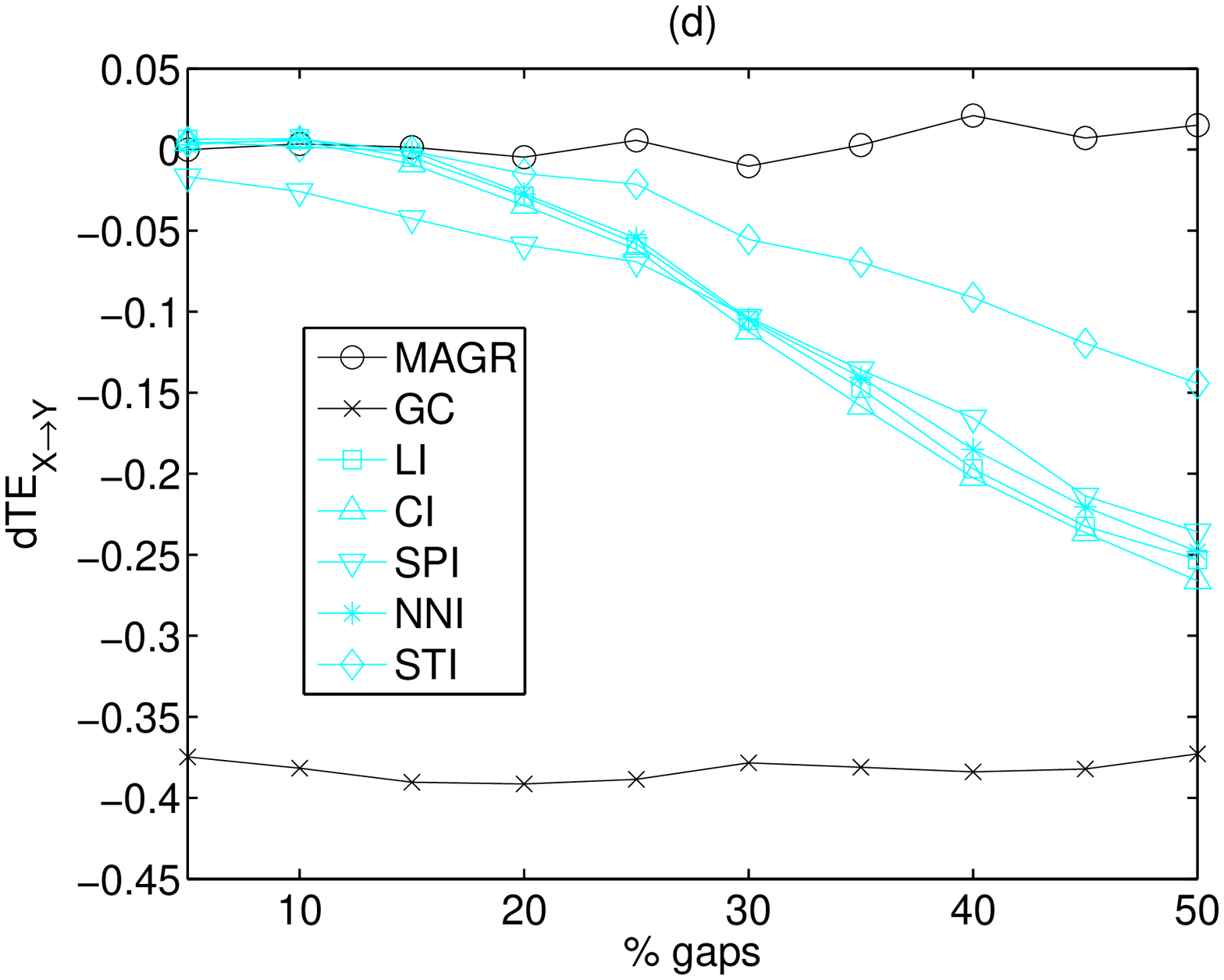}}}
\caption{Difference in the transfer entropy dTE$_{X\rightarrow Y}$
as a function of the percentage of gaps in time series of length $N=1500$, for the gap-treating techniques as shown in the legend. (a) MVAR system and $m=1$, (b) coupled Henon map and $m=1$, (c) MVAR system and $m=2$, (d) coupled Henon map and $m=2$.}
\label{fig3}
\end{figure}

The performance of the gap treating techniques for increasing
number of gaps in the time series when using transfer entropy (TE)
is similar to that for the correlation measures in
Sec.~\ref{subsec:resultsccmi}. TE is a more complicated measure
than the correlation measures as it involves the probability
distribution of reconstructed points at higher dimensions, and it
is therefore more data demanding. In order to obtain sensible TE
estimates we use in the simulations time series of length
$N=1500$. In Fig.~\ref{fig3}, the results on dTE$_{X\rightarrow
Y}$ are shown for the MVAR and coupled Henon systems and for the
two smallest embedding dimensions, $m=1$ and $m=2$.
Again MAGR gives dTE at the zero level for any $g$ up to 50\% of
$N$. Gap-filling techniques achieve this only for very small $g$,
and as $g$ increases dTE approaches the largest in magnitude dTE obtained by GC
(for any $g$). This pattern holds for both the linear and the
nonlinear system and for $m=1$ and $m=2$. For the coupled Henon
map and $m=2$ (Fig.~\ref{fig3}d), the gap-filling techniques
improve the TE estimation for small $g$ and all but SPI give dTE
at the zero level. A possible explanation for this is that for the
Henon system the information of the one missing sample is
partially compensated by the next existing sample, as the
individual dynamics are two dimensional.

\begin{figure}[htb]
\hbox{\centerline{\includegraphics[width=7cm]{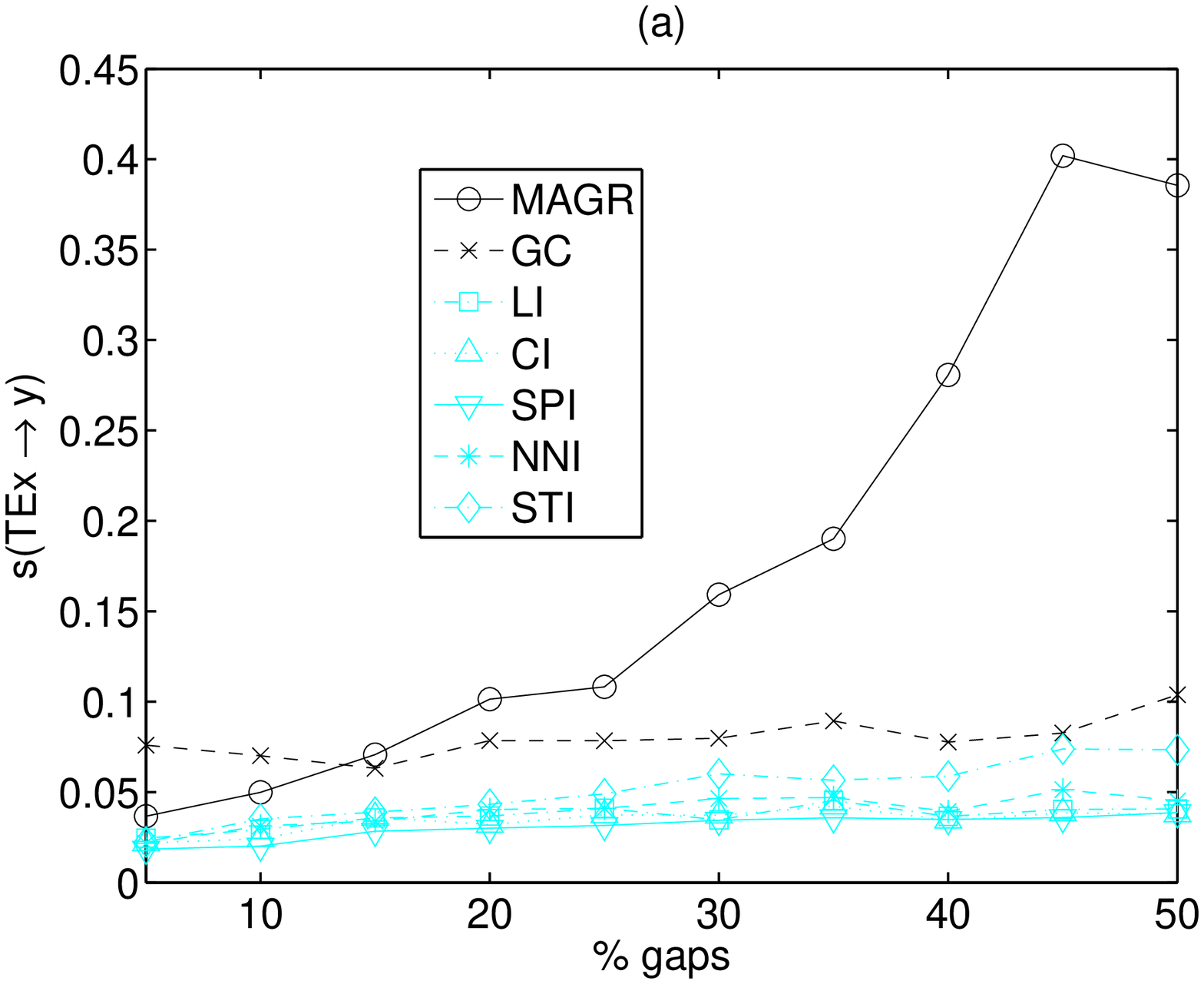}
\includegraphics[width=7cm]{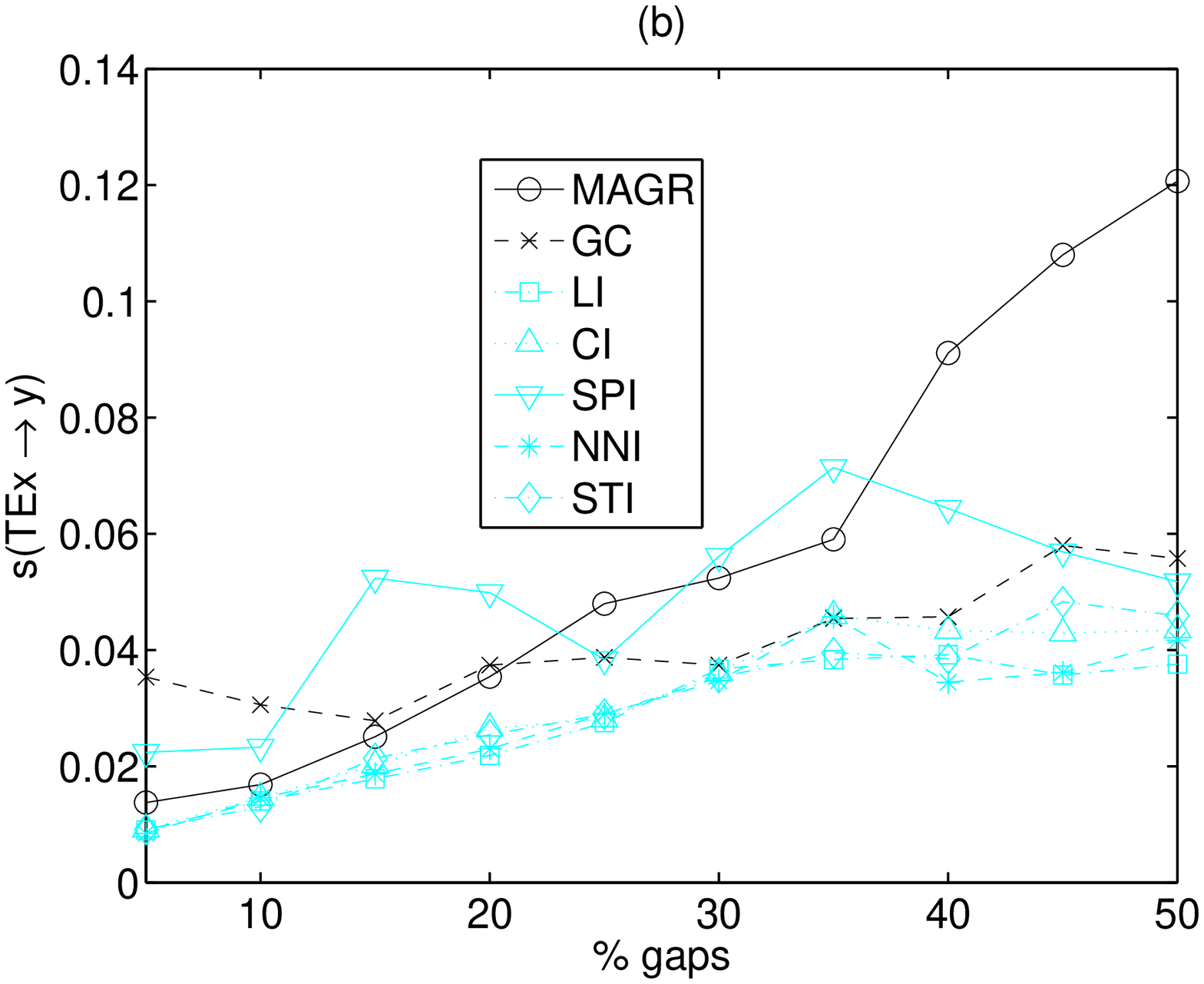}}}
\caption{Standard deviation of the transfer entropy TE$_{X\rightarrow Y}$, $s(\mbox{TE}_{X\rightarrow Y})$,
as a function of the percentage of gaps in time series of length $N=1500$, for the gap-treating techniques as shown in the legend. (a) MVAR system and $m=2$, (b) coupled Henon map and $m=2$.}
\label{fig4}
\end{figure}

We stress here that the reported results are for the mean dTE over
50 realizations. Though MAGR gives mean dTE at the zero level the
variance of dTE increases with the number of gaps because the 
effective number of data points is reduced (see Fig.~\ref{fig4}). 
For example,
for $m=2$, $N=1500$ and $g=750$, the joint data matrix of
$(y_{t+1},x_t,x_{t-1},y_t,y_{t-1})$ is drastically reduced from
1498 to less than 150 points left to compute the correlation sums
in (\ref{eq:TE2}). For the MVAR
system the estimation variance is not affected by the gap percentage for all
but MAGR methods, whereas for MAGR it increases steadily with the gap percentage (Fig.~\ref{fig4}a). 
For the coupled Henon maps, the increase of the standard deviation of TE with MAGR 
is smaller than for the MVAR system (Fig.~\ref{fig4}b). Moreover, for the coupled Henon map, the standard
deviation of TE increases with the gap percentage up to 35\% for the other methods as well at 
a similar rate to MAGR, and then stabilizes, whereas for MAGR it 
continues to increase. It seems that the increase of the variance of the TE estimation
with the gap percentage varies with the underlying system and method parameters. 
A possible remedy for stabilizing the
TE estimation would be to use larger $r$ for smaller time series
(in the simulations $r$ is fixed to 0.2). We do not consider this
as a drawback of MAGR but merely as an indication of the
insufficiency of TE estimation when there are far too many gaps in
the time series, especially when $m$ gets large. In fact, using
any of the gap-filling techniques the TE estimation has the same
variance as for the original time series irrespective of $g$, but
always fails to match the expected TE as if there were no gaps,
and the deviation of dTE increases with $g$.

\subsection{The effect of time series length}
\label{subsec:length}
\par
\begin{figure}[htb]
\hbox{\centerline{\includegraphics[width=7cm]{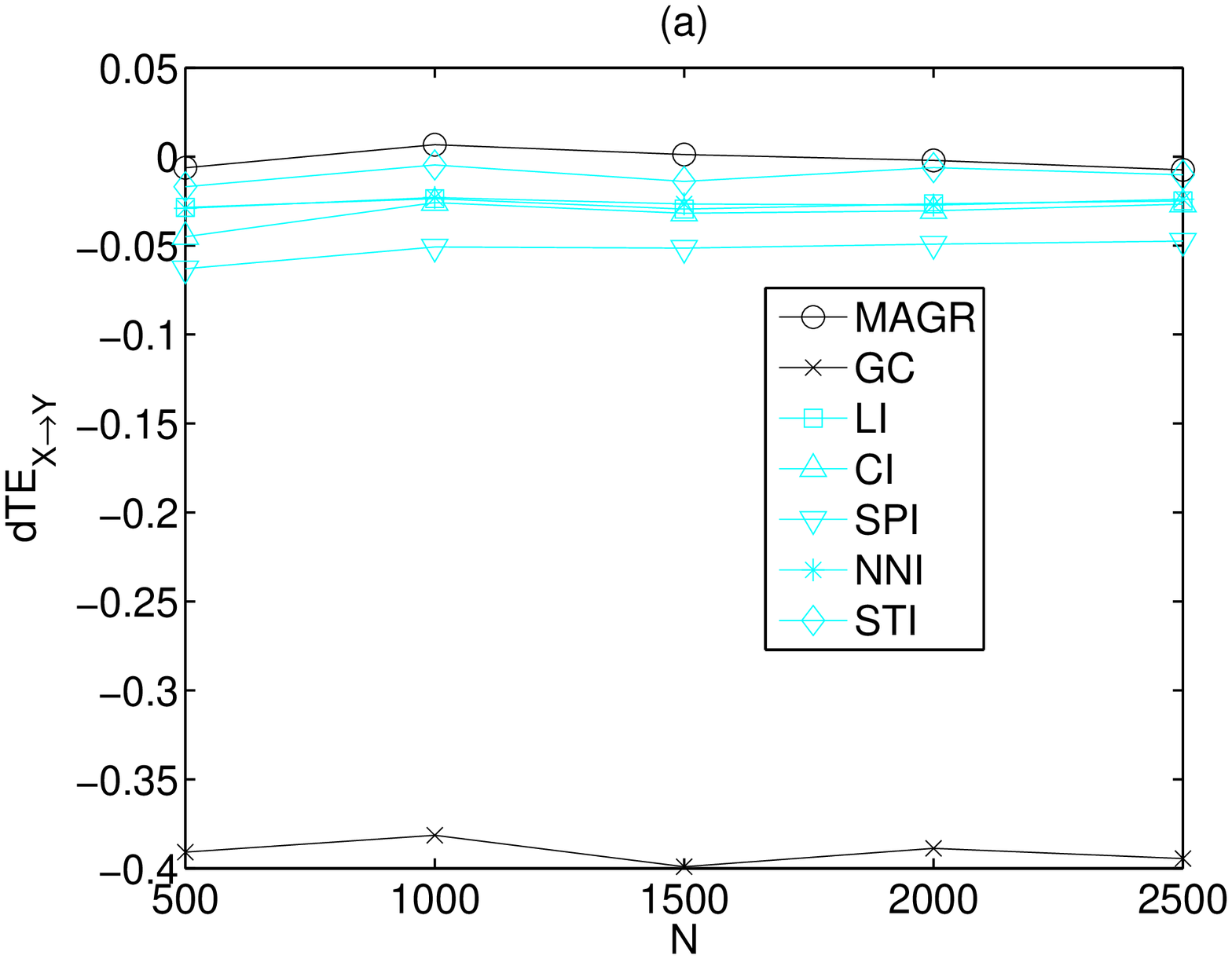}
\includegraphics[width=7cm]{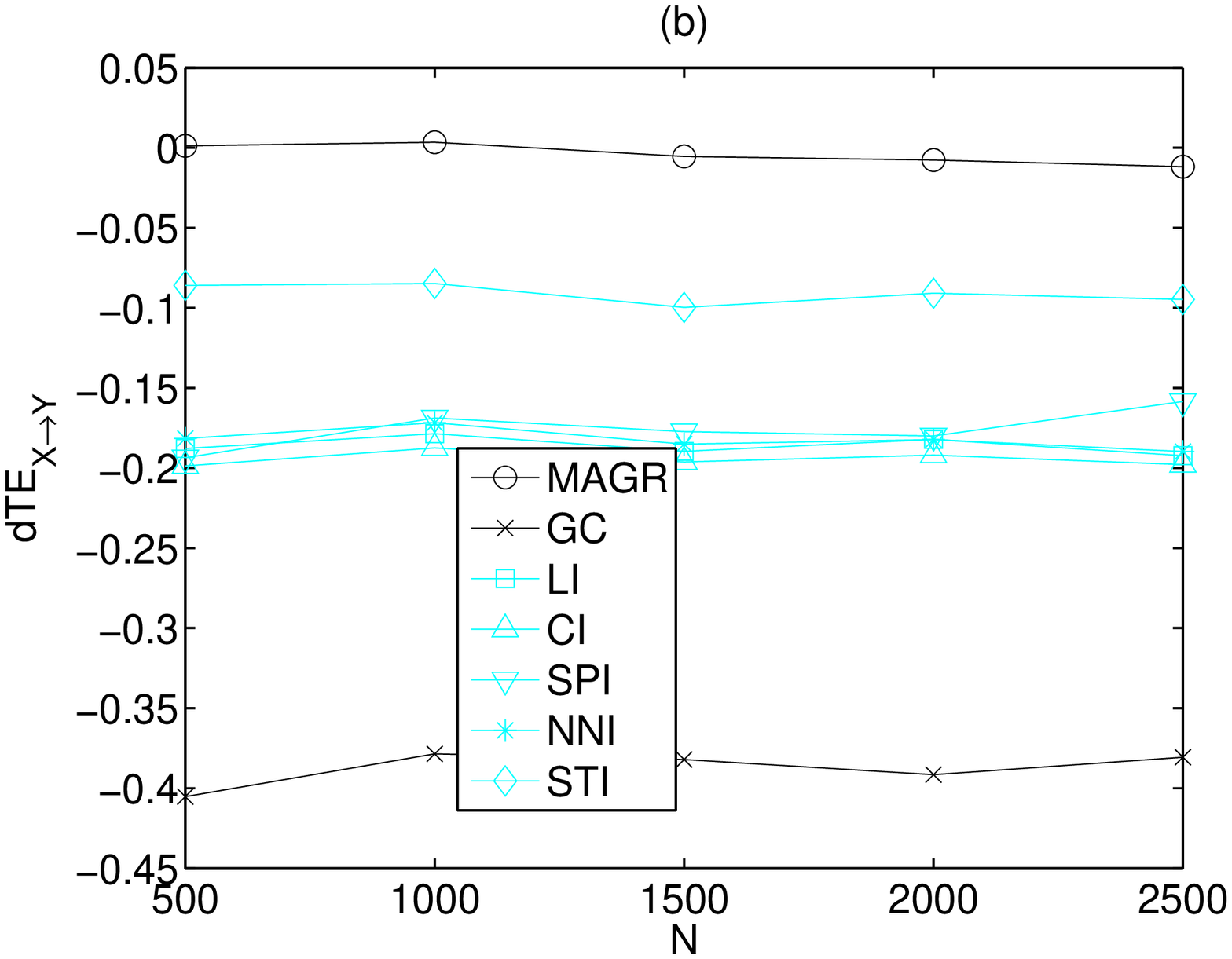}}}
\caption{The dependence of dTE$_{X\rightarrow Y}$ ($m = 2$) for
the gap treating methods, as shown in the legend, on the length
$N$ of the time series generated by the coupled Henon map. The
percentage of gaps is 20\% in (a) and 40\% in (b).}
\label{fig5}
\end{figure}

In the simulations above we fixed the length $N$ of the generated
time series and varied the number of gaps in it in order to assess
the dependence of the performance of the gap treating techniques
on the density of gaps. Here we want to examine the dependence on
$N$, and therefore we fix the percentage of gaps to 20\% and to
40\%, and vary $N$ from 500 to 2500. We concentrate on TE as it is
the most data demanding of the three studied measures and also set
$m = 2$. The results for the coupled Henon map are shown in
Fig.~\ref{fig5}.
Similar were the results for the MVAR system (not shown). It is
clear that all methods are stable with respect to $N$ giving the
same dTE for the whole range of tested $N$. Again dTE obtained
by GC gives the bound of worst performance, while MAGR performs
best attaining always dTE at the zero level. On the other hand,
the gap-filling methods give dTE close to zero (but larger in
magnitude than the dTE from MAGR) for low gap percentage
(Fig.~\ref{fig5}a), but for larger gap percentage dTE deviates
away from zero (Fig.~\ref{fig5}b).

\subsection{The effect of block gaps}
\label{subsec:blocks}

In real time series, there may be consecutive missing values
termed block gap. Here, we examine how the gap treating techniques
cope with the presence of block gaps in the time series. We
consider both fixed size blocks and varying size blocks. To
generate time series with block gaps, we remove block of elements
of a given size from the time series so that the final number of
missing values is $g$. Also we apply the restriction of not having
connected or overlapping block gaps.

\subsubsection{Fixed size block gaps}

The results we obtained for block gaps of fixed size were essentially the same
as for the single gaps. Again the two bounds of performance were
set by GC, giving the largest deviation of the connectivity
measure from the measure on the non-gappy time series, and MAGR,
giving good matching of the connectivity measure on the gappy and
non-gappy time series. The gap-filling techniques exhibited again
a deviation increasing with the number of block gaps. The only
difference as compared to the case of single gaps is that STI was
significantly improved, giving much less deviation than the other
gap-filling methods.

\begin{figure}[htb]
\hbox{\centerline{\includegraphics[width=7cm]{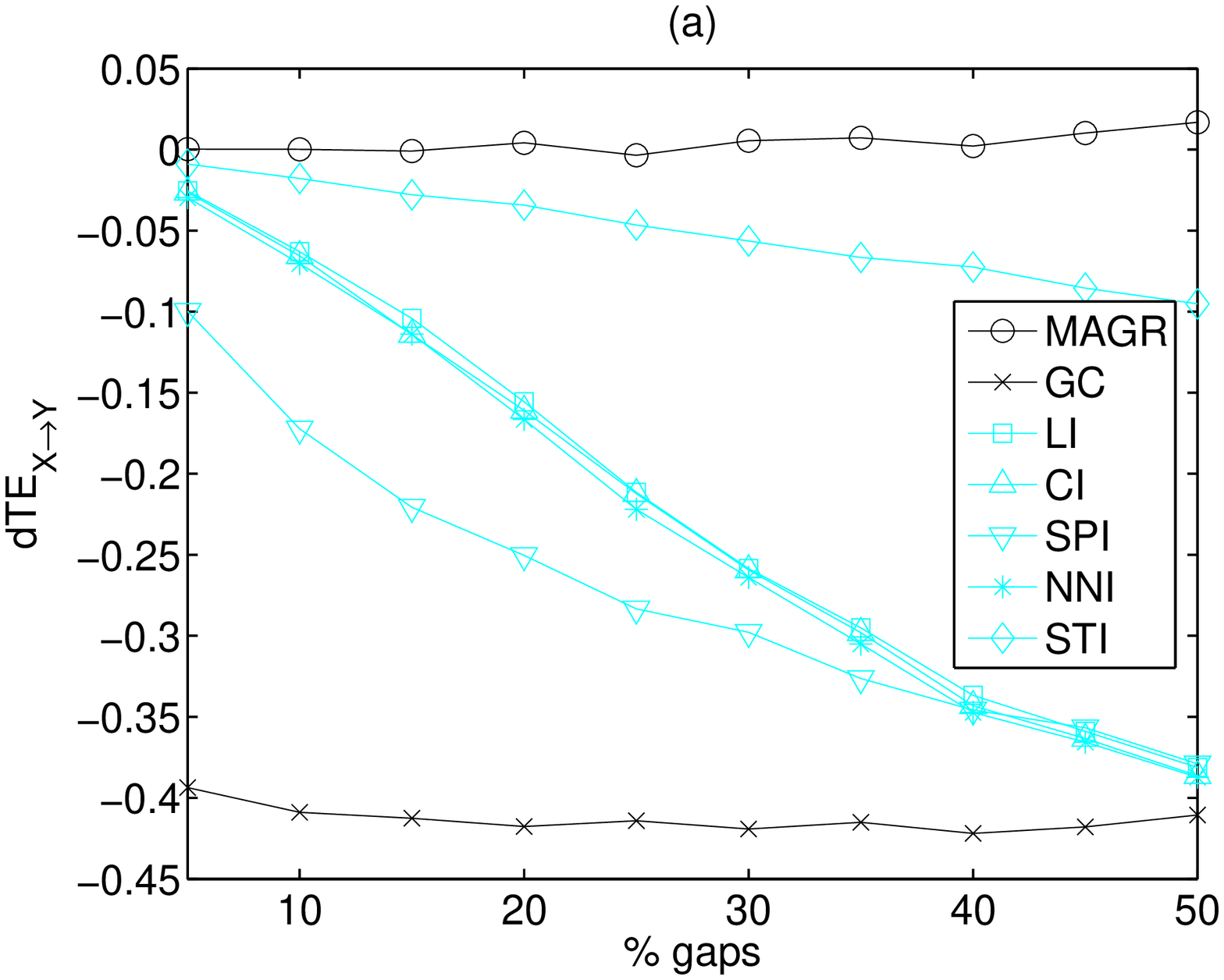}
\includegraphics[width=7cm]{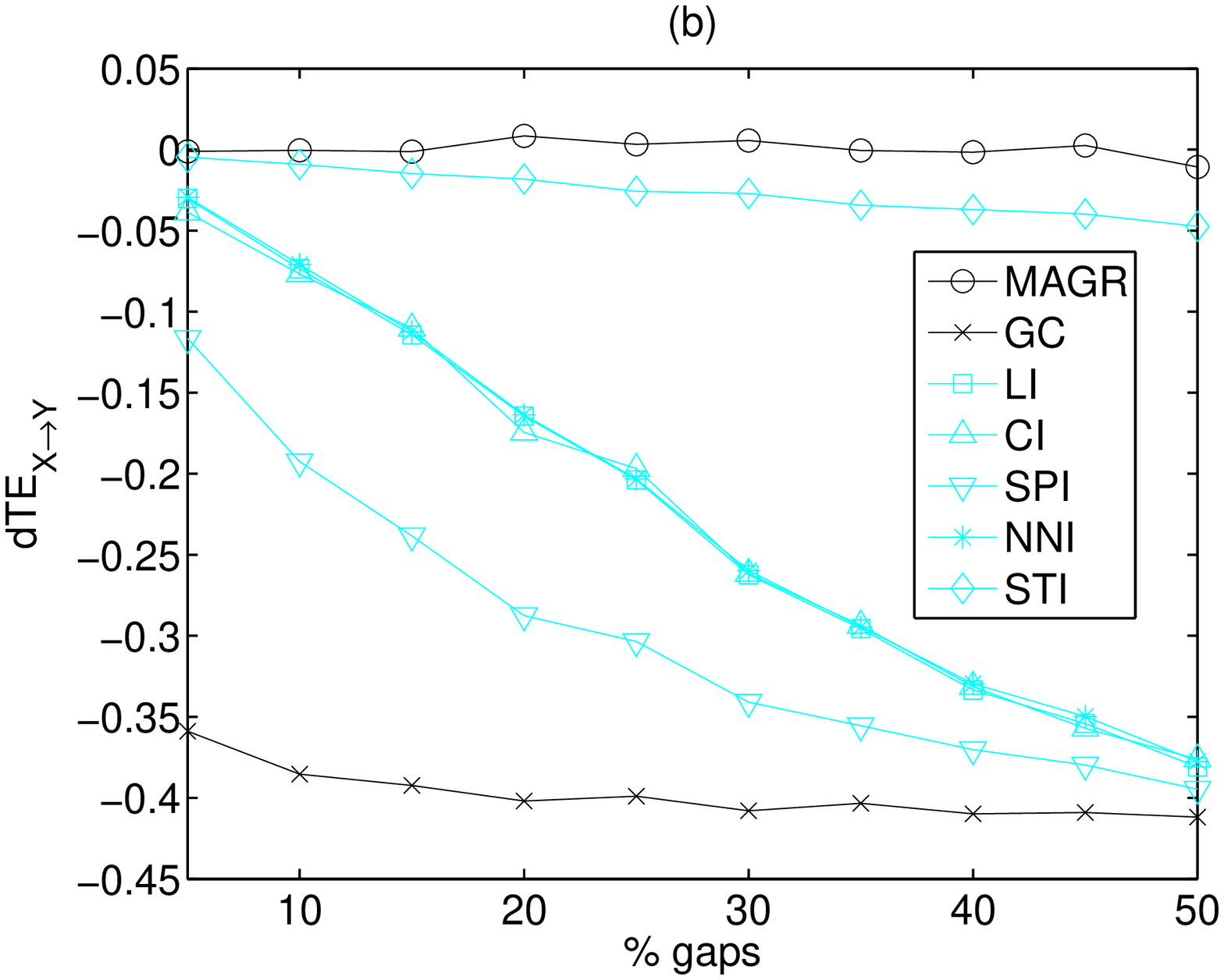}}}
\hbox{\centerline{\includegraphics[width=7cm]{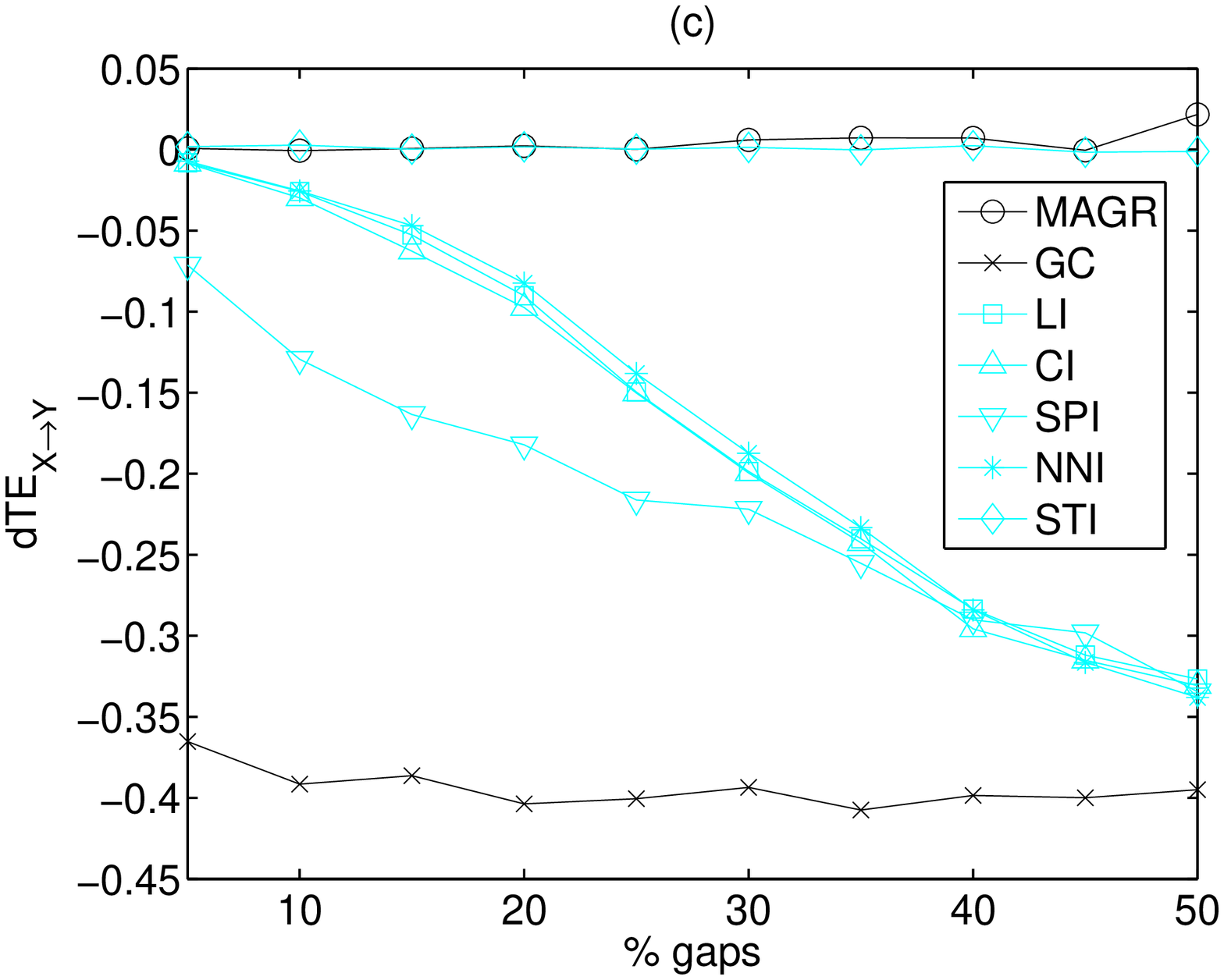}
\includegraphics[width=7cm]{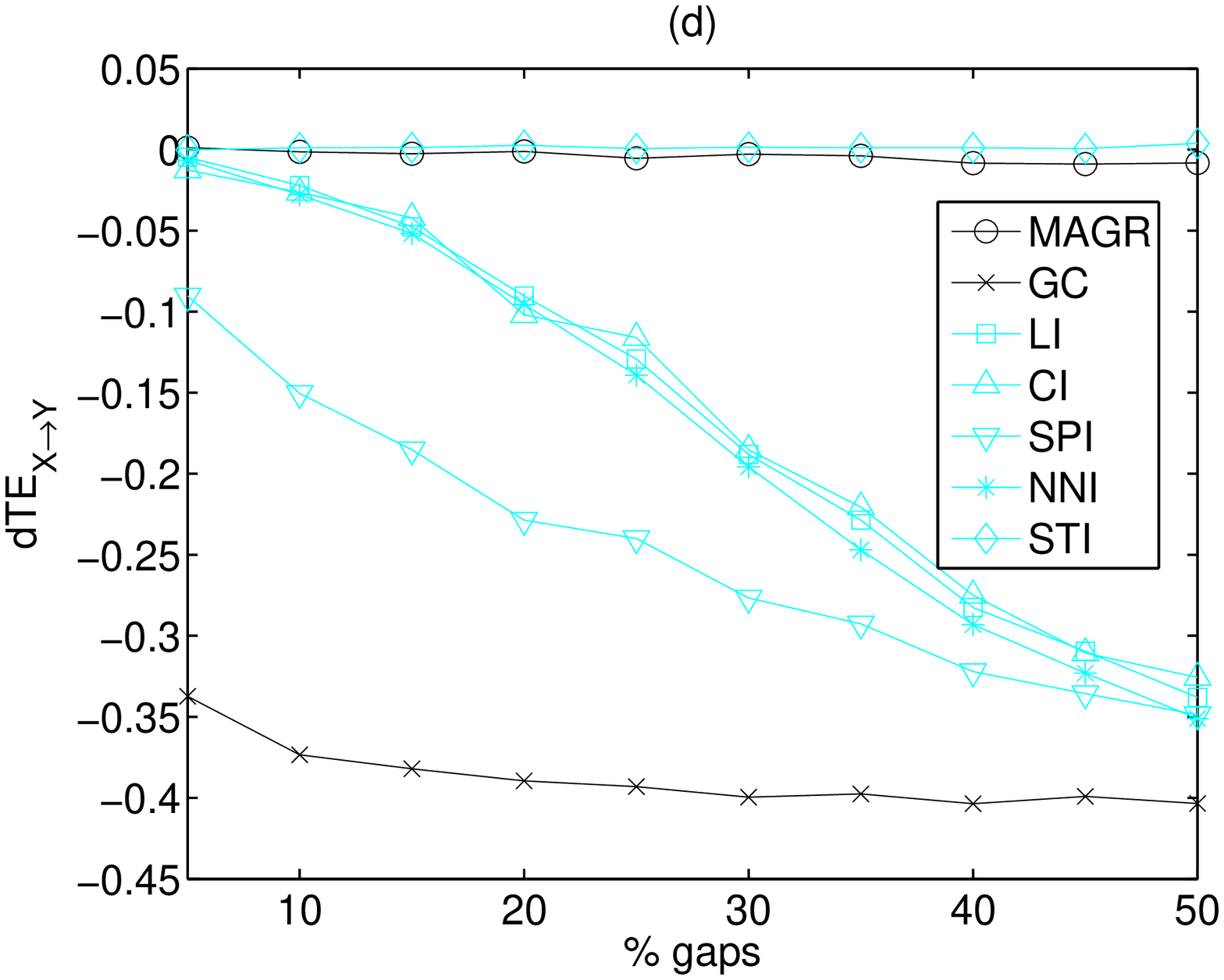}}}
\caption{Difference in the transfer entropy dTE$_{X\rightarrow Y}$
as a function of the percentage of gaps in time series of length $N=1500$ from the coupled Henon map for the gap-treating techniques as shown in the legend. (a) $m=1$ and block gaps of size 5, (b) $m=1$ and block gaps of size 10, (c) $m=2$ and block gaps of size 5, (d) $m=2$ and block gaps of size 10.}
\label{fig6}
\end{figure}

In Fig.~\ref{fig6}, the results of the simulations are shown for
the coupled Henon map, $N=1500$, and TE estimation for $m=1$ and
$m=2$ using block gaps of size 5 and 10.
The results confirm the performance of GC and MAGR as discussed
above and show that STI distinguishes from the other four
gap-filling methods. In particular for $m=1$, dTE obtained by STI
has the smallest deviation from the zero level over all
gap-filling methods, increasing also with $g$, but slower when the
size of the block gaps gets larger, while the other gap-filling
methods do not seem to be affected by the size of the block gap.
For $m=2$, STI improves further and dTE stays at the zero level as
for MAGR. A possible explanation for this is that because STI does
not make any assumption on the underlying dynamics and fills the
gaps in a stochastic manner, it turns out to be more suitable when
many consecutive values are missing than assuming deterministic
dynamics as done by the other gap-filling methods.

With regard to the variance of the TE estimation when using MAGR,
we note that the increase of variance with $g$ is much slower in
the case of fixed size block gaps because fewer points of the full joint data
matrix are dropped than when the gaps are of single type. For
example for the setup of Fig.~\ref{fig6}d, for $g=750$ the number
of points in the joint data matrix falls from 1498 to about 350
points after applying MAGR, to be compared to 150 points for
single gaps. This has a direct effect on the standard deviation of
TE estimation obtained with MAGR in the 50 realizations, e.g. for
$g=750$ this is 0.05 for block size 10, which is much smaller
than for single gaps.

\subsubsection{Varying size block gaps}

We relaxed the constrain of the fixed size of the block gap and
allowed for variable block gap size ranging from 1 to 15, drawn
randomly, i.e. from a discrete uniform distribution in the range
1 to 15. As before, when generating the gappy time series the
restriction of not having consecutive or overlapping block gaps
was applied, and blocks were added until the predefined gap
percentage was reached.

\begin{figure}[htb]
\hbox{\centerline{\includegraphics[width=7cm]{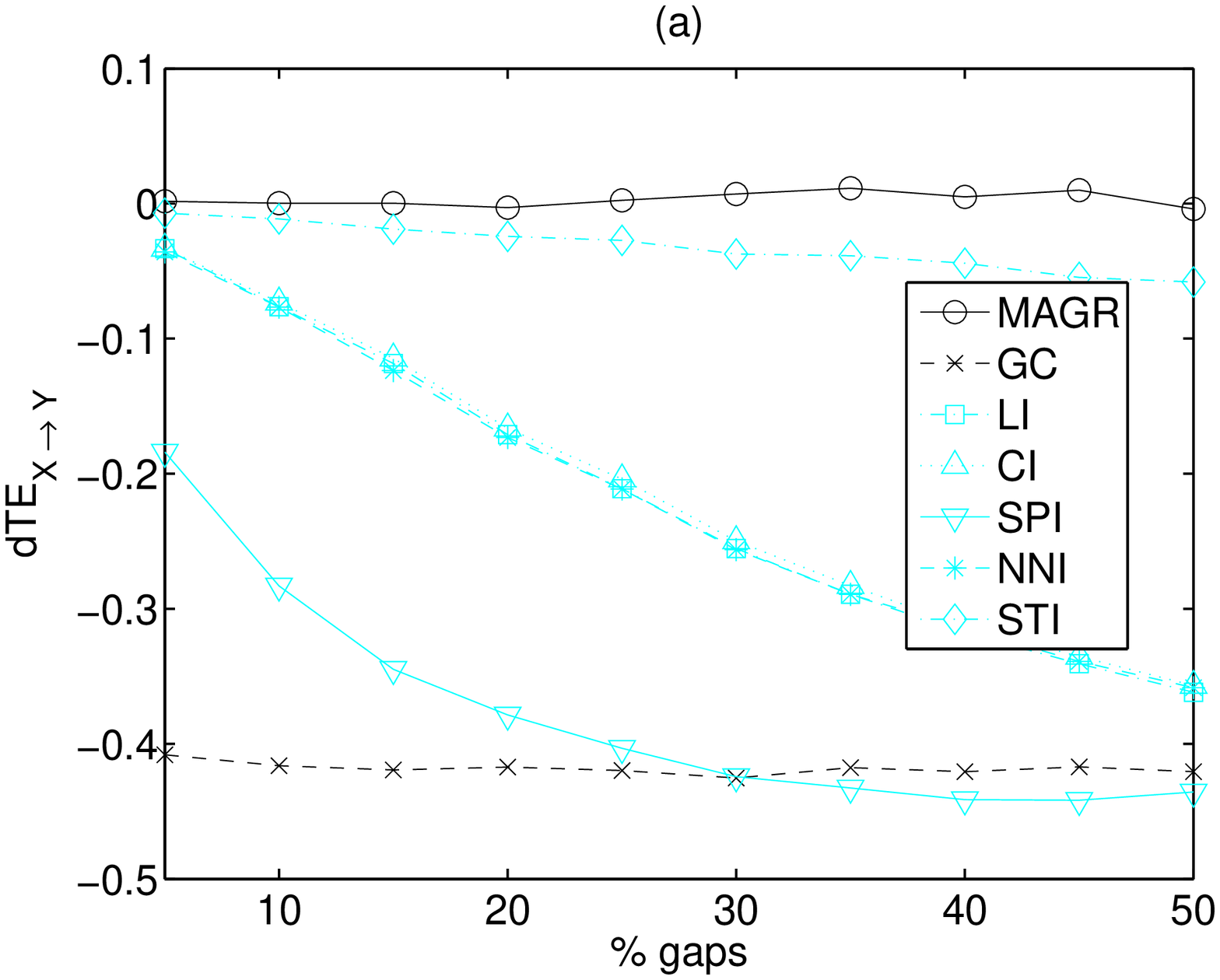}
\includegraphics[width=7cm]{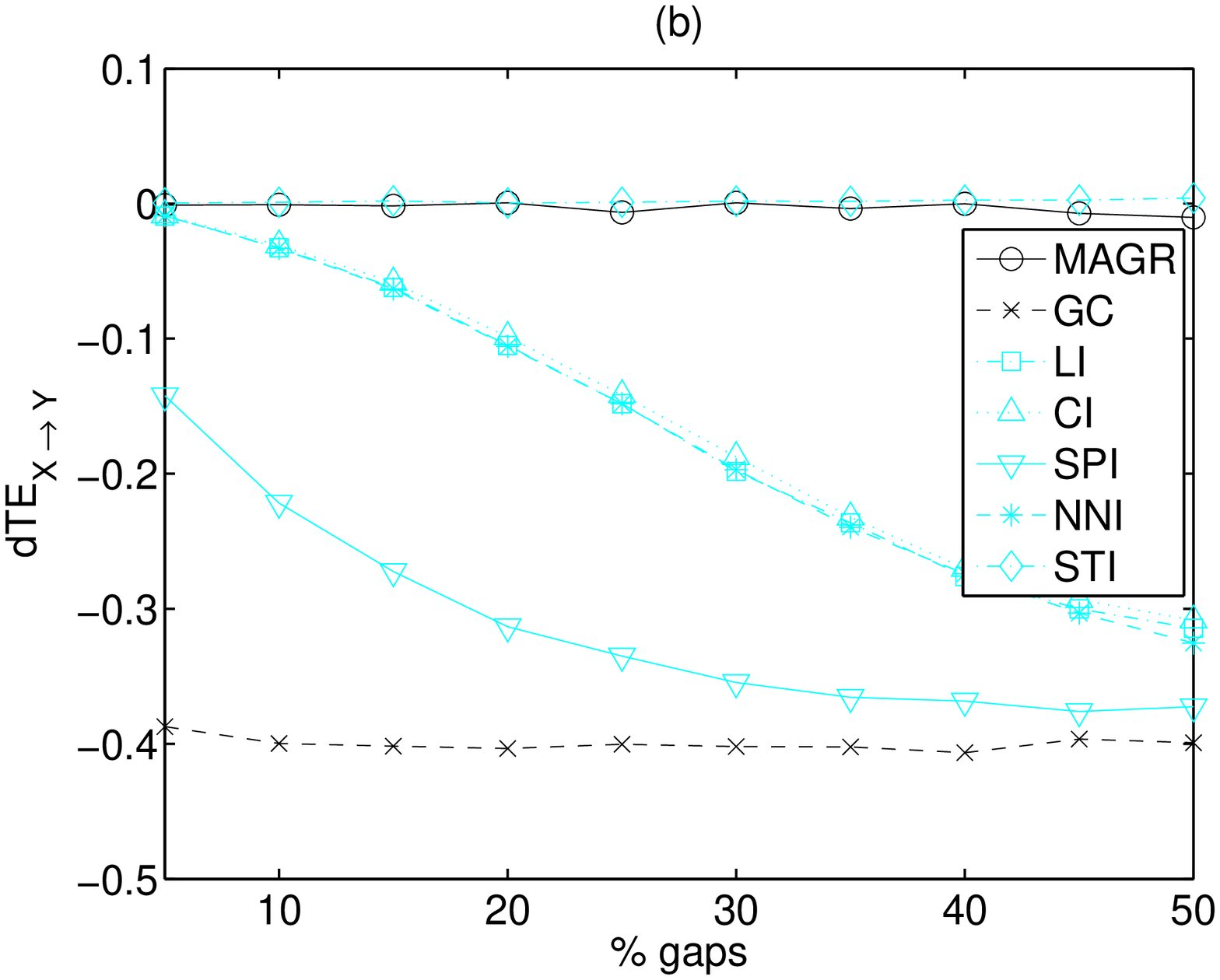}}}
\caption{Difference in the transfer entropy dTE$_{X\rightarrow Y}$
as a function of the percentage of varying size gaps in time series of length $N=1500$ 
from the coupled Henon map, for the gap-treating techniques as shown in the legend.
(a) $m=1$ and (b) $m=2$.}
\label{fig7}
\end{figure}

The results follow a similar pattern as in the case of the block
gaps of fixed size, as shown in Fig.~\ref{fig7} on the same
experimental setting for $m=1$ and $m=2$.
The best performing method is MAGR with second best STI, which deviates from dTE's zero level slowly with the gap percentage for $m=1$ but stays at the zero level as MAGR for $m=2$. The performance of the other methods is inferior with GC being for all gap percentages at the same large deviation and the rest declining towards this with the gap percentage. Interestingly, for $m=1$ the CI method approaches and even exceeds for large gap percentage the lower bound of GC. 

\section{Application}
\label{sec:Application}
\par

Here we assess MAGR in conjunction with the correlation and causality 
measures on financial time series. We used daily data of five European stock market
indexes from 13 October 2008 to 8 September 2011\footnote{The data were downloaded from
finance.yahoo.com.}. 
The selected period covers the financial crisis starting at about
September 2008 as well as the sovereign debt crisis of 2010-2011, including turbulent as well as less turbulent periods. Moreover, we observed in this period the least occurrences of gaps with a rate of 2.6\% to 3.7\% of the data, which is much lower than for other past periods. Also, the length of the selected period is adequately large for the estimation of the connectivity measures when we artificially insert gaps.  

\begin{figure}[htb]
\hbox{\centerline{\includegraphics[width=7cm]{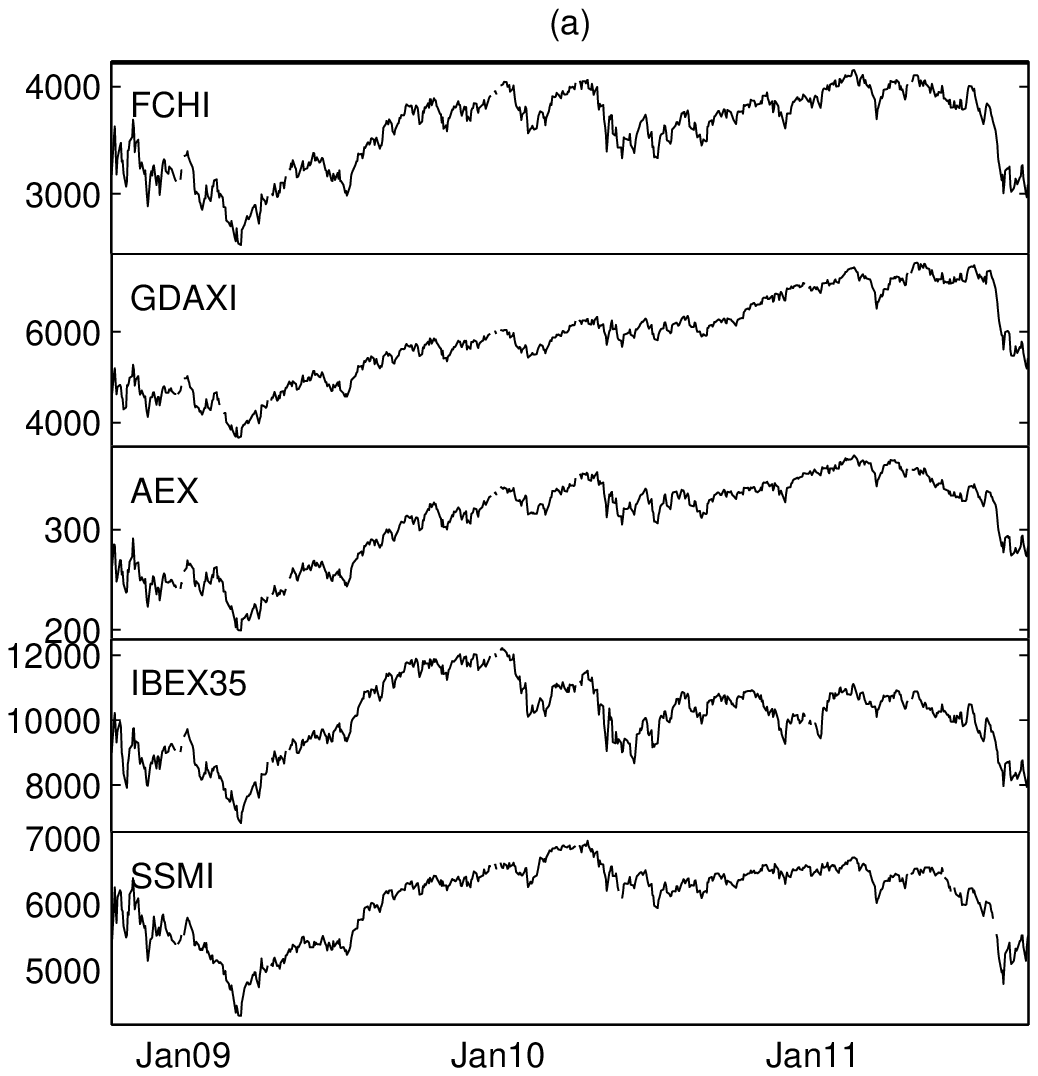}
\includegraphics[width=7cm]{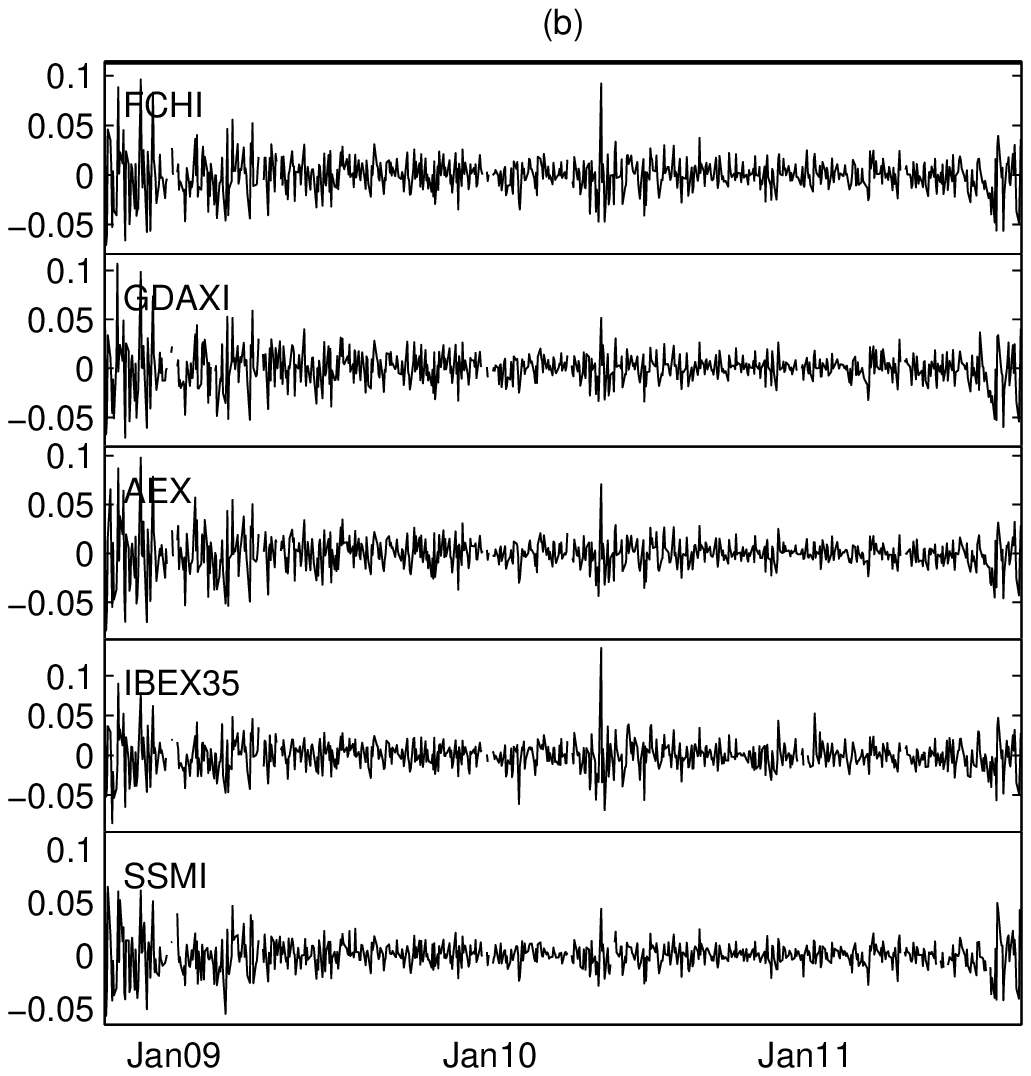}}}
\caption{(a) The close index of 5 financial markets, as indicated in the panels, in the period 13 Oct 2008 to 8 Sep 2011. (b) The returns of the indexes in (a).}
\label{fig:financeseries}
\end{figure}

The stock markets selected include big and medium economies, countries that experienced the financial turmoil in different degrees of severity and also countries that share a common currency as well as others that they have their own. The final sample consists of the stock market of France (FCHI), Germany (GDAXI), Netherlands (AEX), Spain (IBEX35) and Switzerland (SSMI). The historical close index prices are shown in Fig.~\ref{fig:financeseries}a and their returns (log differences) in Fig.~\ref{fig:financeseries}b.  
Despite the increased volatility in the starting period we consider the returns time series of the five markets as fairly stationary and proceed with the analysis on them.  

The analysis consists of the estimation of cross correlation, cross mutual information and TE on the original time series and the time series added with random gaps. The latter were constructed by inserting blocks of gaps of random size (ranging from 1 to 5) into the original time series thus achieving an overall level of missing values of 10\% and 20\%.

\begin{figure}[htb]
\hbox{\centerline{\includegraphics[width=7cm]{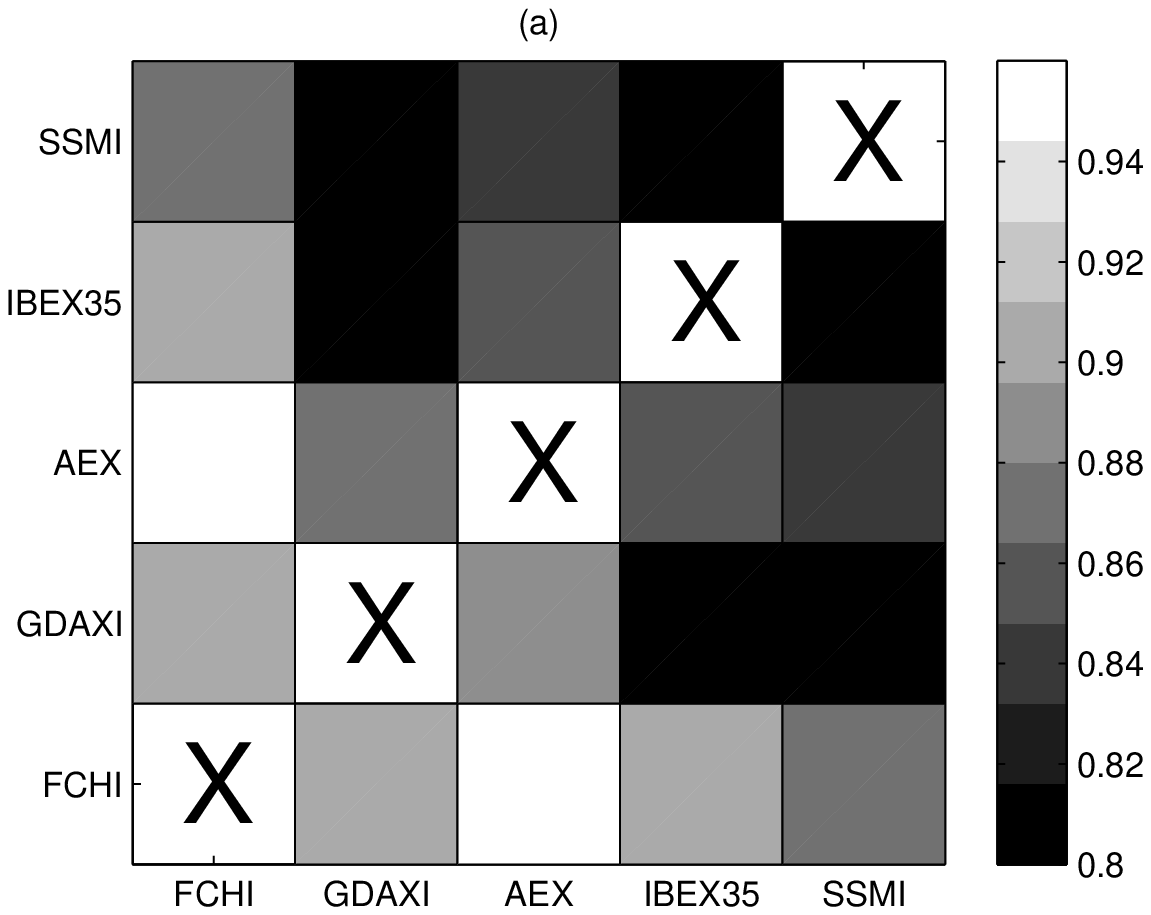}
\includegraphics[width=7cm]{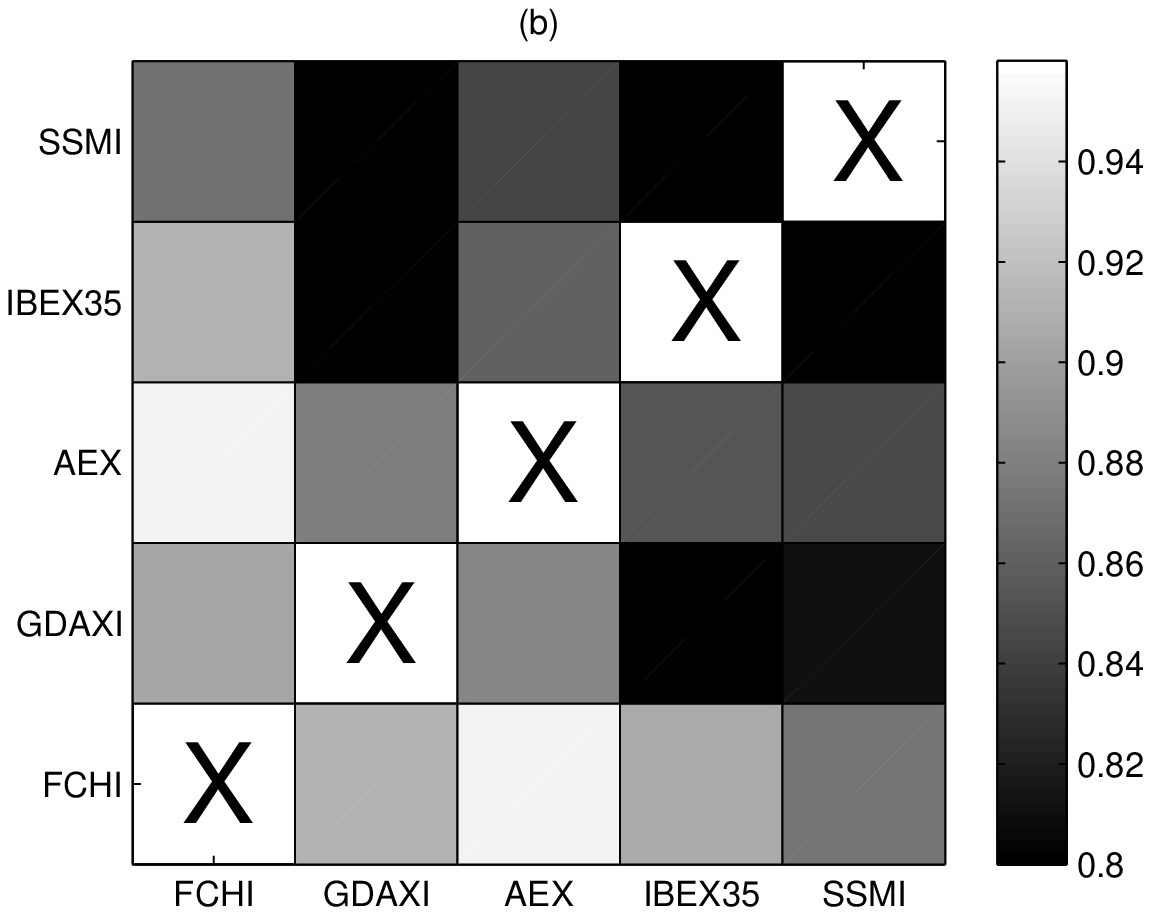}}}
\hbox{\centerline{\includegraphics[width=7cm]{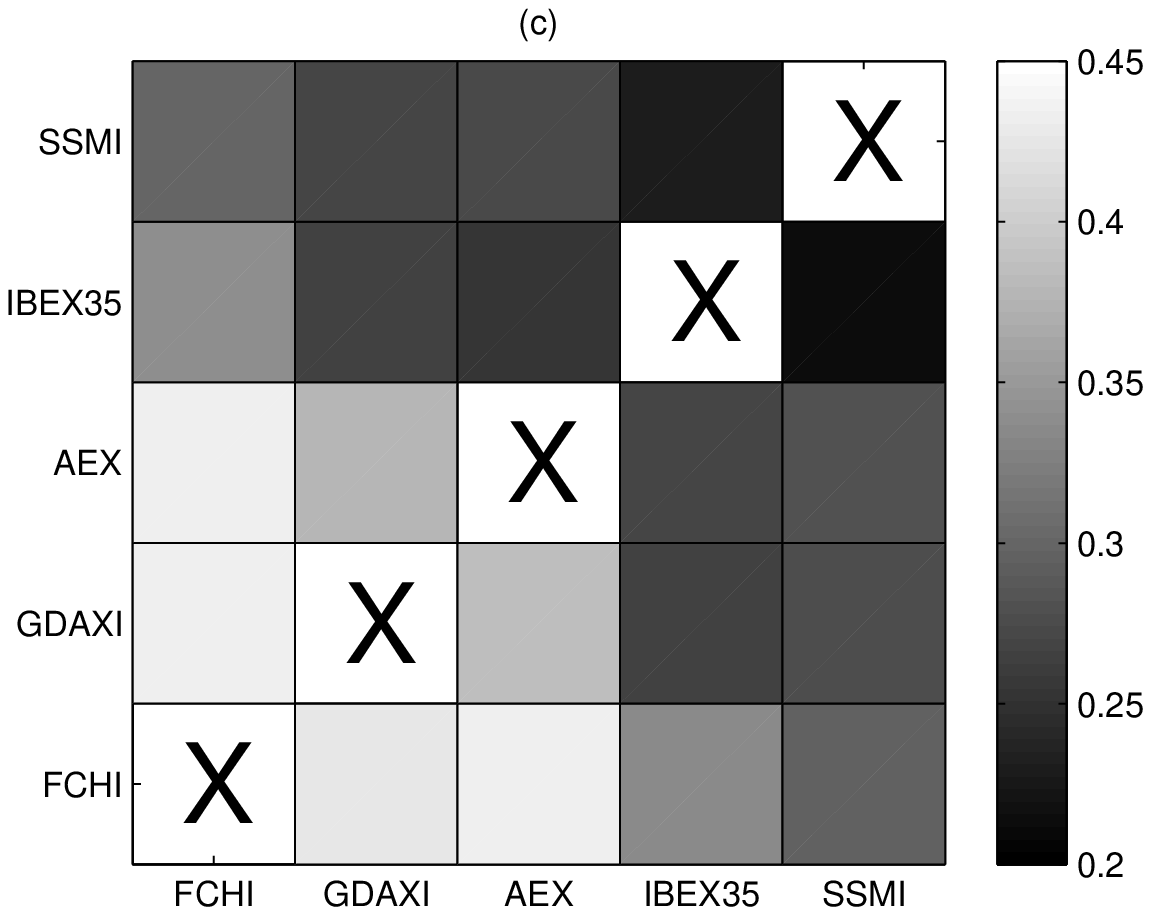}
\includegraphics[width=7cm]{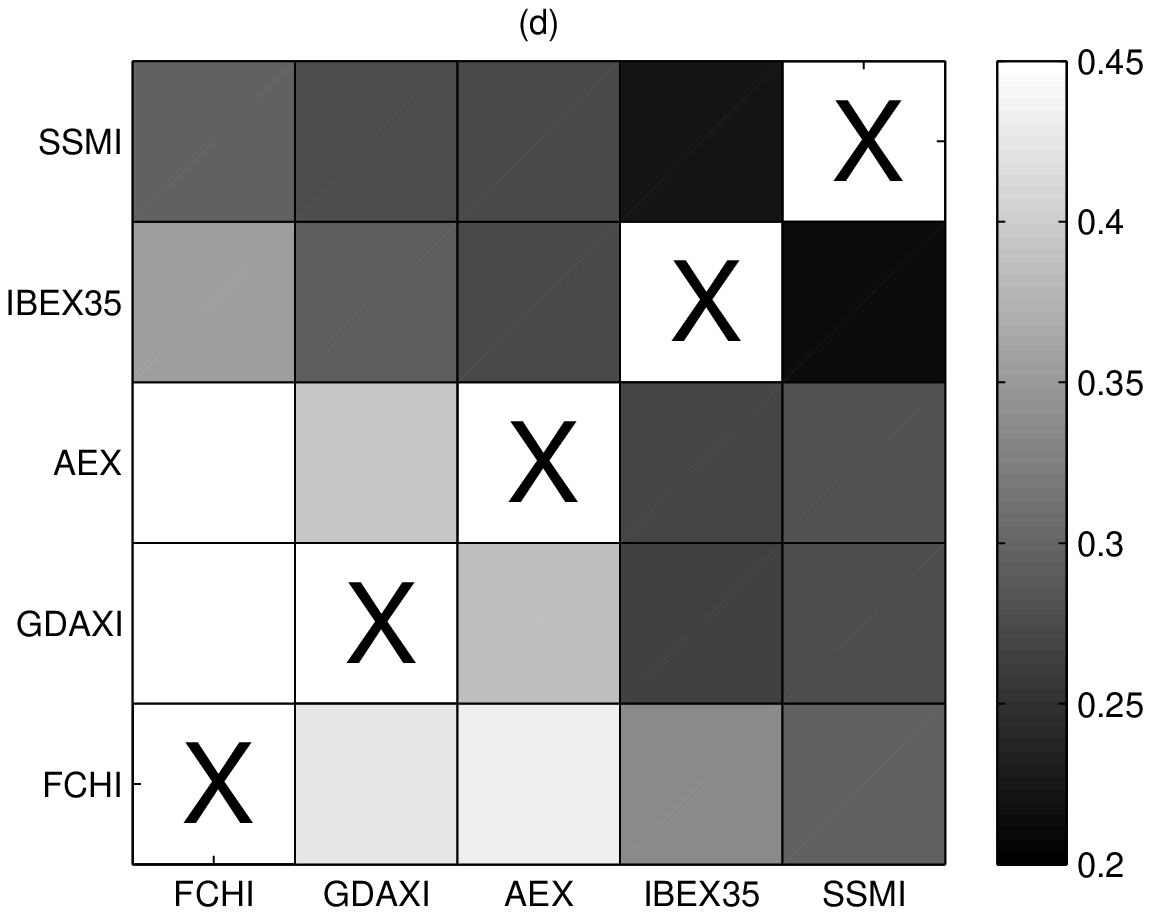}}}
\caption{Color maps at grey scale of the correlation measures on the original time series of the six financial indexes at the lower-right triangular part and the gappy time series (using MAGR) at the upper-left triangular part. (a) cross-correlation and 10\% gaps, (b) cross-correlation and 20\% gaps, (c) mutual information and 10\% gaps, (d) mutual information and 20\% gaps.}
\label{fig:financecormut}
\end{figure}

The estimated cross correlation and cross mutual information for the gappy time series using MAGR is very close to the respective values obtained on the original time series, as shown in Fig.~\ref{fig:financecormut}. 
For each of the panels in Fig.~\ref{fig:financecormut}, each cell of the upper-left triangular part for the gappy time series has the same gray-scale color as the symmetric cell in the lower-right triangular part for the original time series, indicating that for each pair of financial indexes the correlation measure has approximately the same value when computed on the original and gappy time series. The only slight deviation that can be observed is for the cross mutual information when the percentage of gaps is 20\% (see the pair GDAXI -– IBEX35 in Fig.~\ref{fig:financecormut}d). Thus the correlation between the financial indices could be reliably estimated even if a good number of values were missing. The results show that AEX, FCHI and GDAXI correlate stronger to each other than IBEX35 and SSMI. 
It is noted that the estimation of the correlation measures was stable with respect to the gaps inserted though the length of the gappy time series was reduced from about 735 to 640 for 10\% gaps and to 490 for 20\%, depending on the financial indexes. Actually,
for the pair GDAXI –- IBEX35 the cross mutual information gave somehow larger value for 20\% gaps (from 0.265 to 0.293), the length of the gappy time series was the smallest being 474.

\begin{figure}[htb]
\hbox{\centerline{\includegraphics[height=5.4cm]{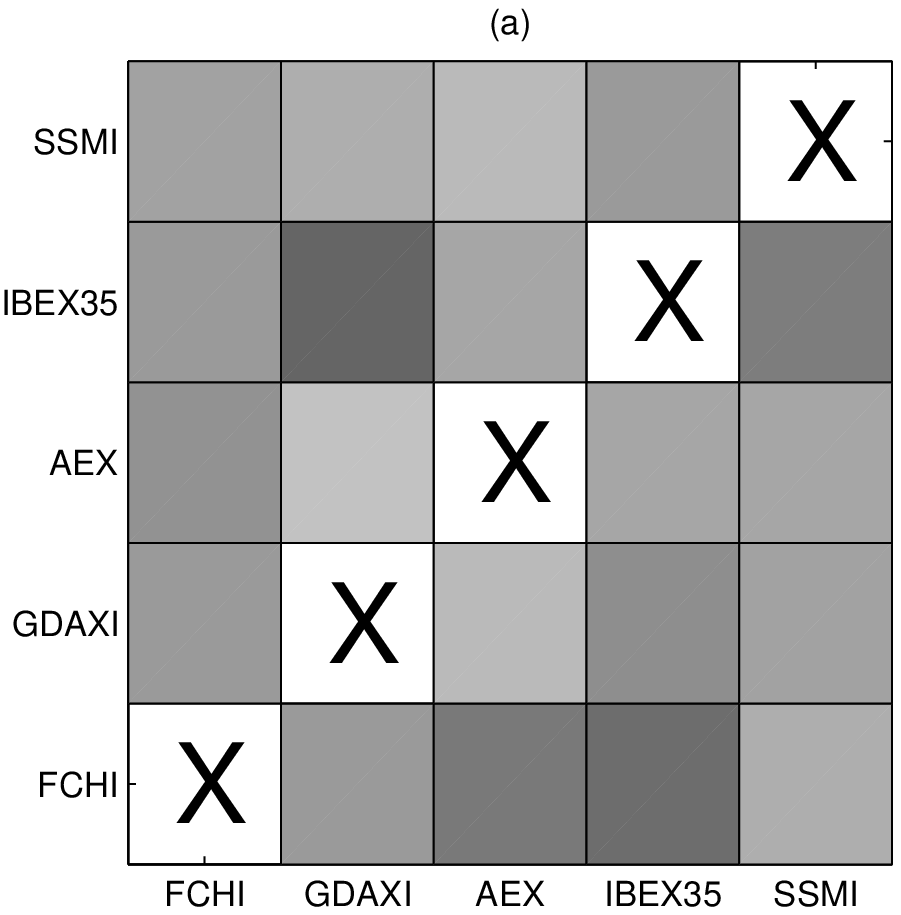}
\includegraphics[height=5.4cm]{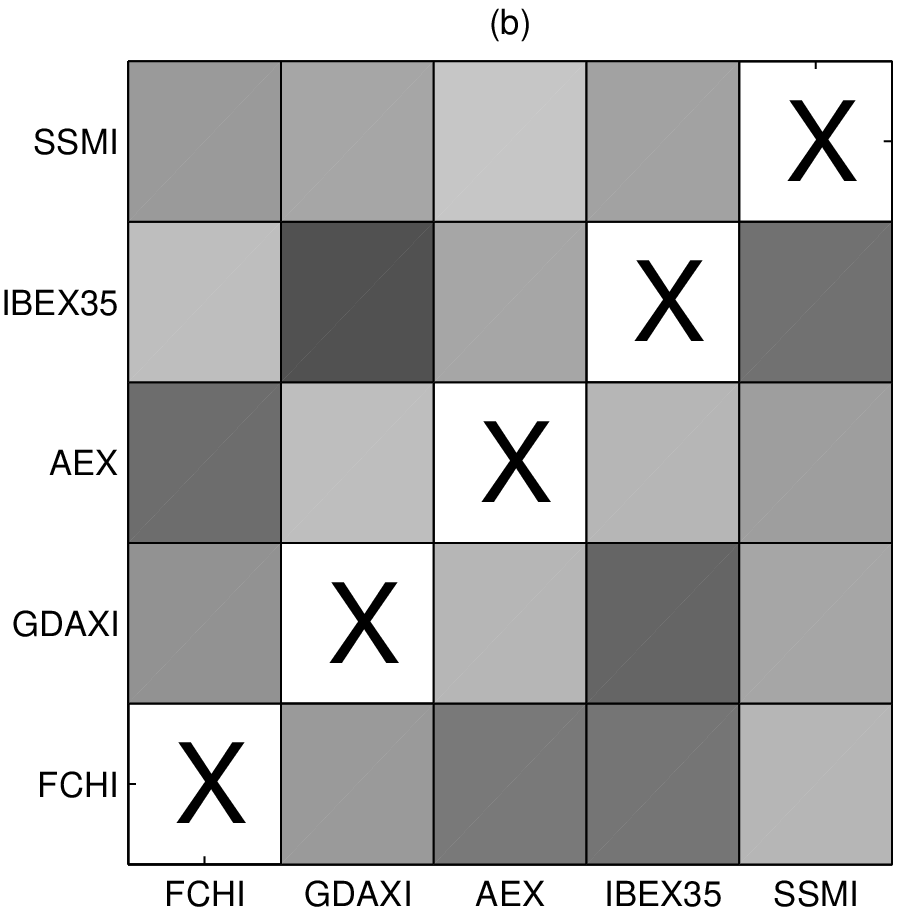}
\includegraphics[height=5.4cm]{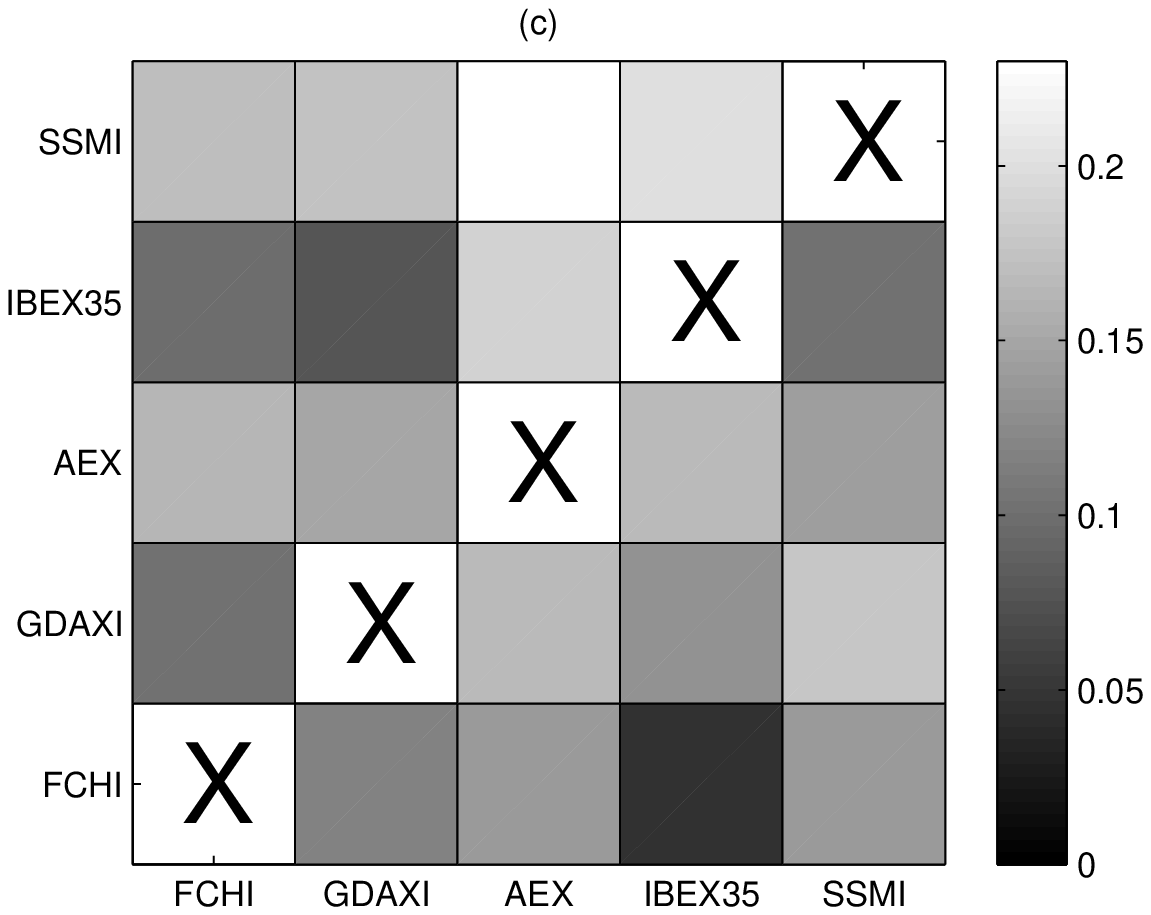}}}
\caption{Color maps at grey scale of the transfer entropy on the six financial indexes: (a) original time series, (b) time series with 10\% gaps, and (c) time series with 20\% gaps. The driving effect is from row to column.}
\label{fig:financete}
\end{figure}

Regarding the transfer entropy, its values on the gappy time series using MAGR deviate more from the original financial time series, as shown in Fig.~\ref{fig:financete}.
Adding 10\% gaps does not change significantly the estimation of TE, and although there are small changes in the TE values, the overall pattern of causal relationships remains essentially the same. Still, there is no systematic bias and the deviations are randomly positive and negative. For 20\%, the original pattern of causal relationships is retained relatively well but there are substantial differences, visible in the color maps, i.e. comparing Fig.~\ref{fig:financete}a to Fig.~\ref{fig:financete}c. Though the variance increases with the gap percentage there is no bias and the TE value decreases, e.g. from FCHI to IBEX35, and increases, e.g. from SSMI to AEX, in a random manner. The
reason for having for 20\% gap percentage larger variance in the TE estimation than in the correlation estimation is that TE is more data demanding. Indeed for the simplest setup of TE used here ($m=1$, $\tau=1$) the effective number of data points is only slightly smaller than for the correlation measures, and specifically the number of data points for the original time series is about 725 and decreases to 620 for 10\% gaps and to 460 for 20\%, depending on the financial indexes. 

\section{Discussion}
\label{sec:Discussion}

We have presented a simple idea for treating the gaps in
multivariate time series when computing a connectivity measure,
such as the cross correlation, cross mutual information and
transfer entropy. The considered connectivity measures suppose a
joint data matrix, where the rows are time ordered and each row
regards a vector of present, and possibly past and future
observations of both variables. The idea is to omit the rows that
contain empty entries, corresponding to missing values, and
proceed with the calculations on the matrix of reduced rows. The
important advantage of this approach, called measure adapted gap
removal (MAGR), is that the dynamics of the (possibly) coupled
system is intact, which is in contrast to all the gap-filling
methods, replacing the gaps under a stochastic or deterministic
model for the underlying dynamics. Certainly, any hypothesized
model cannot be appropriate for every problem. Our simulations on a
linear and a nonlinear coupled system revealed the inadequacy of the
gap-filling methods (we considered linear, cubic, spline, nearest
neighbor and stochastic interpolation), and confirmed the
appropriateness of MAGR, in estimating connectivity measures in
bivariate time series with gaps.

An apparent disadvantage of MAGR is that the amount of available
data for the computation of the connectivity measure is reduced at
a degree that depends on the number of gaps $g$ in conjunction
with the type of gaps (the same $g$ can regard many single missing
values or few groups of consecutive missing values called block
gaps), and the parameters of the connectivity measure, e.g. the
embedding dimension $m$ in the measure of transfer entropy. Thus
using MAGR in a setting involving, say, a large $g$ and $m$, there
may be insufficient data points to compute the connectivity
measure and the estimate may have large variance. Actually, this
is equivalent of having an equally small non-gappy time series,
and the simulations showed indeed the good matching of the two
estimates of the connectivity measure, on the gappy time series
treated by MAGR and the non-gappy time series of equivalent
length. The simulations showed that MAGR gives no deviation of the
average connectivity measure estimated on the gappy and non-gappy
time series, irrespective of $g$, but the estimate may have larger
variance (implied by the amount of available data), whereas the
gap-filling methods give deviation that increases with $g$. The
worst performance was observed for the gap closure method, which
is actually based on the strongest, and generally least valid,
assumption of independent observations. Thus for time series with
many gaps, the estimate of a connectivity measure using a
gap-filling method will tend to deviate, whereas using MAGR it
will be at the level of the estimation on the time series of
reduced length.

We also found that when the same number of gaps occur in blocks of
fixed or varying size, it makes no difference for the performance of
the gap-filling methods, except for the stochastic interpolation,
which improved with the size of the block gap. On the other hand,
for MAGR the estimation is equally well matched but now the
variance is decreased as the rows of empty entries in the joint
data matrix is reduced. The latter does not regard measures that
involve only present values in the rows of the joint data matrix,
such as the zero lag cross correlation and cross mutual
information, for which the data reduction in MAGR is the same for
single and block gaps.

The above findings suggest that, depending on the sensitivity of
the connectivity measure estimate to the time series length,
estimates on different bivariate time series of the same length
but with varying amount of gaps should not be directly compared,
either using a gap-filling method or MAGR. Using MAGR, the
comparison should be done on the basis of joint data matrices with
non-empty entries of the same size. This situation may be met in
many applications using windowing of multivariate time series
records where gaps of irregular size and frequency may occur, e.g.
in the analysis of multi-channel electroencephalograms (EEG) or
financial market indices. However, our exemplary study on a set of six financial market indices showed that MAGR can provide reliable estimates of correlation and causality measures even when there are significant gaps in the time series. Besides fluctuations in the estimation of the transfer entropy that increased with the gap percentage, the causal relationships for each pair of indexes could be robustly estimated. It is noted
however, that these results were obtained with a simple setup for the transfer entropy, and increasing the embedding dimension would result in drastically decrease of the effective data size using MAGR and subsequently larger variance in the estimation of transfer entropy. Still the obtained evidence from the simulations and the application in finance shows that for small percentage of gaps or large time series MAGR is a suitable approach to obtain reliable estimates of cross-correlation and Granger causality.

In this work, we demonstrated the MAGR approach on some bivariate
connectivity measures. Though MAGR is measure specific and has to
be adapted to the selected measure, it is straightforward to adapt
it to any measure that has as a core a joint data matrix, which
encompasses most of the measures on multivariate time series. For
example, MAGR can directly be applied to partial transfer entropy,
conditioning the driving-response effect on other observed
variables \cite{Vakorin09,Papana12}. Simply, the joint data matrix
is expanded containing columns for the other observed variables.
Certainly, if there are asynchronous gaps in the time series of
the other variables, the size of the matrix will be further
reduced affecting the stability of the measure estimation. 
The systematic analysis using multivariate causality measures
is the subject of a forthcoming study.
MAGR
can also be applied to models of univariate and multivariate time
series, as they also rely on a joint data matrix. For example, for
an univariate autoregressive model, the matrix of lagged variables
has the role of the joint data matrix, and after the elimination
of the rows of empty entries the ordinary least squares can be
applied to compute the model parameters. The extension to
multivariate autoregressive models is straightforward.

\bibliographystyle{elsarticle-harv}


\end{document}